\documentclass[aps,10pt,prd,notitlepage,showpacs,nofootinbib,superscriptaddress,compressed]{revtex4-1}

\pdfoutput=1 
\usepackage[utf8]{inputenc}
\usepackage[english]{babel}
\usepackage{amsmath}
\usepackage{graphicx}
\usepackage{dcolumn}
\usepackage{pbox}
\usepackage{amssymb}
\usepackage{epsfig}
\usepackage[dvipsnames]{xcolor}
\usepackage{slashed}
\usepackage{amssymb}
\usepackage{ mathrsfs }
\usepackage{color}
\usepackage[font=small]{caption}
\usepackage[font=small]{subcaption}
\usepackage{url}
\usepackage{braket}
\usepackage{hyperref}
\usepackage{fontawesome5}
\usepackage{booktabs}
\graphicspath{{pic/}}

\definecolor{DarkBlue}{rgb}{0.7, 0.4, 1} 
\definecolor{Blue}{rgb}{0, 0.8, 0} 
\definecolor{MyLightBlue}{rgb}{0.5,0.7,1.9}
\definecolor{MyGreen}{rgb}{0.0,0.2, 0.0}
\definecolor{MyBrickRed}{rgb}{0, 0.5, 0.2}
\RequirePackage{hyperref}
\hypersetup{colorlinks, citecolor=Blue,linkcolor=DarkBlue, urlcolor=Green}

\newcommand{\bea}{\begin{eqnarray}}
\newcommand{\eea}{\end{eqnarray}}
\newcommand\dd{\mathrm{d}}

\makeatletter

\makeatletter
\renewcommand\@makecaption[2]{%
  \par
  \vskip\abovecaptionskip
  \begingroup
  
   \small\rmfamily
    \begingroup
     \samepage
     \flushing
     \let\footnote\@footnotemark@gobble
     \@make@capt@title{#1}{#2}\par
    \endgroup
  \endgroup
  \vskip\belowcaptionskip
}
\makeatother

\DeclareUnicodeCharacter{2212}{-}
\newcommand {\black} {\color{black}}

\def\21{\mathrm{$SU(2)_L \otimes U(1)_Y$}}

\begin{document}
\title{Right handed neutrino production from $Z^\prime$ interactions in forward search experiments}
\author{ShivaSankar K.A}\email{a-shiva@particle.sci.hokudai.ac.jp}
\affiliation{Department of Physics, Hokkaido University, Sapporo 060-0810, Japan}
\author{Souvik Das}\email{souvik\_das@tamu.edu}
\affiliation{Department of Physics and Astronomy, Mitchell Institute for Fundamental Physics and Astronomy, Texas A\&M University, College Station, Texas 77843, USA}
\affiliation{Department of Physical Sciences, Indian Institute of Science Education and Research Kolkata, Mohanpur, 741246, India}
\author{Arindam Das}
\email{adas@particle.sci.hokudai.ac.jp}
\affiliation{Institute for the Advancement of Higher Education, Hokkaido University, Sapporo 060-0817, Japan}
\affiliation{Department of Physics, Hokkaido University, Sapporo 060-0810, Japan} 
\author{Sanjoy Mandal}
\email{smandal@kias.re.kr}
\affiliation{Korea Institute for Advanced Study, Seoul 02455, Korea} 
\begin{abstract}
We study two general $U(1)$ extensions of the Standard Model (SM) those generate tiny neutrino masses via the seesaw mechanism after  general $U(1)$ breaking. These models predict a new neutral gauge boson ($Z'$) and right-handed neutrinos (RHNs), the latter introduced for anomaly cancellation and neutrino mass generation. In both scenarios, left- and right-handed fermions couple differently to the $Z'$, and RHNs mix with light neutrinos, enabling variety of decay modes. Focusing on the high-luminosity LHC (HL-LHC) and the future FASER2 experiment, we explore RHN pair production from $Z'$ decays in two cases: (i) long-lived $Z'$ decays to visible modes and long-lived RHNs, and (ii) short-lived $Z'$ decays to long-lived RHNs, which further decay visibly inside FASER2. We estimate projected limits on the general $U(1)$ gauge coupling, $Z'$ mass, RHN mass, and light--heavy neutrino mixing for various $U(1)$ charge assignments, and compare them with current experimental bounds. \href{https://github.com/SouvikPhD/RHN-Detection-with-FASER-2-}{\faGithub}
\end{abstract}
\maketitle

\section{Introduction}
\label{sec:intro}

The experimental evidence of the existence of tiny neutrino mass and flavor mixing \cite{ParticleDataGroup:2020ssz} are long standing puzzles allowing to think about scenarios beyond the Standard Model (BSM) to explain the origin of these puzzles. Among many interesting aspects, a simple and minimal extension of the SM was proposed in \cite{Minkowski:1977sc,Yanagida:1979as,Gell-Mann:1979vob,Mohapatra:1979ia} where SM-singlet Heavy Neutral Leptons (HNLs)/ Right Handed Neutrinos (RHNs) are introduced to naturally explain tiny neutrino masses through the seesaw mechanism. 

Apart from the particle extensions of the SM, another interesting way to explain neutrino masses is to extend the SM minimally by a general $U(1)$ gauge group which introduce a new heavy $Z'$ vector boson. This $U(1)$ symmetry can be spontaneously broken by the vacuum expectation value (VEV) of a scalar field that is a singlet under the SM gauge group but charged under the $U(1)$ gauge group. Three generations of SM-singlet RHNs, charged under the $U(1)$ gauge group, are required to cancel gauge anomalies. Addition of RHNs also facilitate the seesaw mechanism for generating light neutrino masses. One of very simple examples of this scenario is a B$-$L (Baryon minus Lepton) extension of the SM as proposed in \cite{Davidson:1978pm,Davidson:1979wr,Marshak:1979fm,Mohapatra:1980qe}. We find that among many possibilities of the $U(1)$ extension, a linear combination of the U(1)$_Y$ and $U(1)_{\rm B-L}$ gauge groups, which we dubbed as $U(1)_X$, is unique because of its chiral nature \cite{Appelquist:2002mw,Coriano:2014mpa,Das:2017flq}. Under this gauge symmetry, left handed and right handed fermions are differently charged and interact differently with $Z^\prime$, however, three generations of RHNs are uniformly charged under the general $U(1)_X$ \cite{Das:2017flq}.
In addition to that, there is another interesting aspect where RHNs are differently charged under the general $U(1)_X$ gauge group \cite{Das:2017deo} making the BSM scenario free from gauge anomalies. We name it as  an `alternative' scenario. In these models $Z^\prime$ mass and the gauge coupling of the general $U(1)_X$ gauge group are free parameters which can be tested at different experiments. Extensive studies on low and high energy aspects of the $U(1)_X$ scenarios have been performed in \cite{Das:2021esm,Asai:2022zxw,Asai:2023mzl,KA:2023dyz,A:2024shl}.

Being charged under the general $U(1)$ gauge group, these RHNs can be produced from $Z^\prime$ being kinematically allowed. These RHNs can manifest a variety of phenomenological scenarios through prompt decay and displaced vertex analyses in collider experiments \cite{Das:2019fee,Chiang:2019ajm,Cvetic:2018elt,Cvetic:2019rms}. The RHNs being light in mass can also be produced from a sufficiently light $Z^\prime$ coming from meson decay, such a scenario these days has become a point of interest in a recently proposed and designed forward physics experiment called ForwArd Search ExpeRiment (FASER) \cite{Feng:2017uoz,Kling:2021fwx} at the LHC site. Pair production of RHN from light $Z^\prime$ in the context of B$-$L scenario has been studied in \cite{Li:2023dbs} proposing limits on the RHN mass and mixing between light and heavy neutrinos. Light RHNs from meson decay under low scale seesaw scenario has already been tested in the context of FASER \cite{Kling:2018wct} showing the search reach on the mass-mixing plane of the heavy neutrinos. A promising approach to searching for light particles involves the long-lived particle (LLP) scenario, where the particle, after being produced in a collider or accelerator, travels a measurable distance before decaying—ideally into visible final states that include charged particles. Apart from FASER, there is also another experiment which is potentially designed from LLP searches in the operating MoEDAL-MAPP \cite{Frank:2019pgk} experiment. There are also some proposed experiments like FACET \cite{Cerci:2021nlb}, MATHUSLA \cite{Chou:2016lxi}, CODEX-b \cite{Gligorov:2017nwh}, and ANUBIS \cite{Hirsch:2020klk} where long-lived RHN searches could have interesting features in future. Theoretically in the seesaw scenario, light RHN production from meson decay has been studied in \cite{Chun:2019nwi} where prospective limits on the mass-mixing plane has been shown comparing with the existing bounds from different experiments. It is important to point out that, in the case of $U(1)_X$ extensions of the SM, these RHNs interact with the SM weak gauge bosons by the mixing between light and heavy neutrinos although the RHNs are SM-singlet. 

In this paper we consider minimal and general $U(1)$ extension \cite{Das:2017flq} of the SM where three SM-singlet RHNs and an SM-singlet BSM scalar is introduced. In this scenario three RHNs are uniformly charged. After solving anomaly cancellation conditions we find left and right handed fermions are differently charged under the $U(1)_X$ gauge group. After the $U(1)_X$ symmetry breaking Majorana mass of the RHNs are generated. Followed by the electroweak symmetry breaking Dirac mass of the neutrinos are generated allowing the evolution of light neutrino mass through the seesaw mechanism. The breaking of $U(1)_X$ symmetry induces the mass of neutral BSM gauge boson $Z^\prime$ which interact differently with the left and right handed fermions of the model manifesting chiral scenario.

We consider another scenario in this paper which is called the `alternative' scenario \cite{Das:2017deo} where two generations of the RHNs have same charged under the $U(1)_X$ scenario and the third one is differently charged under the $U(1)_X$ scenario. After solving the anomaly cancellation we find that left and right handed fermions are differently charged under the $U(1)_X$ gauge group manifesting the chiral scenario through the interaction with $Z^\prime$. In this scenario, being protected by the $U(1)_X$ symmetry RHNs interact with with BSM Higgs doublet. After $U(1)_X$ and electroweak symmetry breaking neutrino mass is generated by by the seesaw mechanism.

We study RHN production from $Z^\prime$ at the ATLAS interaction point and detection at the FASER2 detector which is the detector upgraded from FASER. At the interaction point, $Z^\prime$ is produced from meson decay and proton bremsstrahlung processes~\cite{Feng:2017vli,FASER:2018eoc,FASER:2018bac}. The decay of $Z^\prime$ can have two possibilities. First, we consider that $Z^\prime$ is long-lived and decays to visible modes after a time interval from its production point. In addition, we also consider the case that $Z^\prime$ is short-lived and decays promptly to pair of RHNs if kinematically allowed. Due to the chiral scenario $Z^\prime$ interacts differently with the left and right handed fermions which affect its production and decay into different modes. Depending on the model and $U(1)_X$ charge decay of $Z^\prime$ into a pair of RHNs could dominate over different modes. This is a very interesting feature of the chiral scenario.

After pair production, RHNs decay into visible modes while both manifesting long-lived nature. As a result, looking at the visible modes of the $Z'$~(RHNs), we constrain the general $U(1)$ gauge coupling with respect to $Z^\prime$ mass for long-lived~(short-lived) $Z^\prime$. Further, for the short-lived $Z'$, we constrain light-heavy mixing with respect to RHN mass. The interactions between the fermions and $Z^\prime$ depend on the general $U(1)$ charges, and then, the production and decay of $Z^\prime$ are also dependent on the charges. As a result, we estimate constrains on the model parameters for each general $U(1)$ charge.

We arrange our paper in the following way. In Sec.~\ref{sec:model}, we discuss the two considered $U(1)$ scenarios including the basic properties of $Z'$ and RHNs. Next in Sec.~\ref{sec:experiment}, we introduce the analyses of the signal events for long-lived and short-lived particle searches in FASER experiment. We provide results and discussions in Sec.~\ref{sec:RD}, and we finally conclude our paper in Sec.~\ref{sec:conc}. 
\section{Model}
\label{sec:model}
We consider general $U(1)$ extensions of the SM where three generations of the RHNs are introduced to cancel all the gauge and mixed gauge-gravitational anomalies. These $U(1)$ extensions of the SM introduce neutral BSM gauge boson $Z'$ which directly interact with the SM fermions and RHNs. In this paper we study two type of general $U(1)$ extensions described as below:
\subsection{Case-I}
The $U(1)_X$ extension of the SM involves three generations of SM-singlet RHNs $(N_R^\alpha)$ which are charged under the $U(1)_X$ gauge group like the SM fermions. Inclusion of thee RHNs cancel the gauge anomalies. In addition, we introduce an SM-singlet scalar field $(\Phi)$ which couples to the RHNs to generate heavy Majorana masses. The particle content of the $U(1)_X$ model is given in Tab.~\ref{tab1} where the corresponding $U(1)_X$ charges of the fermion $f$ are given by $\tilde{x}_f$. Using the conditions of the gauge anomaly cancellation, we can relate these general $U(1)_X$ charges as follows:
\begin{align}
{\rm U}(1)_X \otimes \left[ {\rm SU}(3)_C \right]^2&\ :&
			2\tilde{x}_q - \tilde{x}_u - \tilde{x}_d &\ =\  0~, \nonumber \\
{\rm U}(1)_X \otimes \left[ {\rm SU}(2)_L \right]^2&\ :&
			3\tilde{x}_q + \tilde{x}_\ell &\ =\  0~, \nonumber \\
{\rm U}(1)_X \otimes \left[ {\rm U}(1)_Y \right]^2&\ :&
			\tilde{x}_q - 8\tilde{x}_u - 2\tilde{x}_d + 3\tilde{x}_\ell - 6\tilde{x}_e &\ =\  0~,  \\
\left[ {\rm U}(1)_X \right]^2 \otimes {\rm U}(1)_Y &\ :&
			{\tilde{x}_q}^2 - {\tilde{2x}_u}^2 + {\tilde{x}_d}^2 - {\tilde{x}_\ell}^2 + {\tilde{x}_e}^2 &\ =\  0~, \nonumber \\
\left[ {\rm U}(1)_X \right]^3&\ :&
			{6\tilde{x}_q}^3 - {3\tilde{x}_u}^3 - {3 \tilde{x}_d}^3 + {2\tilde{x}_\ell}^3 - {\tilde{x}_\nu}^3 - {\tilde{x}_e}^3 &\ =\  0~, \nonumber \\
{\rm U}(1)_X \otimes \left[ {\rm grav.} \right]^2&\ :&
			6\tilde{x}_q - 3\tilde{x}_u - 3 \tilde{x}_d + 2\tilde{x}_\ell - \tilde{x}_\nu - \tilde{x}_e &\ =\  0~. \nonumber 
\label{anom-f-1}
\end{align}
Applying the $\mathcal{G}_{\rm SM} \otimes$ $U(1)_X$ gauge symmetry, we can write the Yukawa interactions between the fermions and scalar fields (SM Higgs doublet $H$ and $\Phi$) as 
\begin{equation}
{\cal L}^{\rm Yukawa} = 
- Y_u^{\alpha \beta} \overline{q_L^\alpha} H u_R^\beta
- Y_d^{\alpha \beta} \overline{q_L^\alpha} \tilde{H} d_R^\beta
- Y_e^{\alpha \beta} \overline{\ell_L^\alpha} \tilde{H} e_R^\beta
- Y_\nu^{\alpha \beta} \overline{\ell_L^\alpha} H N_R^\beta-\frac{1}{2} Y_N^\alpha \Phi \overline{(N_R^\alpha)^c} N_R^\alpha + {\rm H.c.}~,
\label{LYk}   
\end{equation}
where $\tilde{H}= i \tau^2 H^*$ with $(\tau^2)$ being the second Pauli matrix.
Applying charge neutrality in each term of Eq.~\ref{LYk}, it is found that the charges of the fermions are related to those of the scalars $H$ and $\Phi$ as follows:
\begin{eqnarray}
-\frac{1}{2} x_H^{} &=& - \tilde{x}_q + \tilde{x}_u \ =\  \tilde{x}_q - \tilde{x}_d \ =\  \tilde{x}_\ell - \tilde{x}_e=\  - \tilde{x}_\ell + \tilde{x}_\nu;\,\, 
2 x_\Phi^{}	=\ - 2 \tilde{x}_\nu~. 
\label{Yuk}
\end{eqnarray} 
Finally, by solving these equations, the charges of the fermions are expressed in terms of $x_\Phi=1$ and $x_H$.
The charge assignments are given in Tab.~\ref{tab1}. 
In this model, all the three generations of each fermion interact with $Z^\prime$ in the same way. 
\begin{table}[t]
\begin{center}
\begin{tabular}{||c|ccc||rcr||}
\hline
\hline
            & SU(3)$_c$ & SU(2)$_L$ & U(1)$_Y$ & \multicolumn{3}{c||}{U(1)$_X$}\\[2pt]
\hline
\hline
&&&&&&\\[-12pt]
$q_L^i$    & {\bf 3}   & {\bf 2}& $\frac{1}{6}$ & $x_q^\prime$ 		& = & $\frac{1}{6}x_H + \frac{1}{3}x_\Phi$  \\[2pt] 
$u_R^i$    & {\bf 3} & {\bf 1}& $\frac{2}{3}$ & $x_u^\prime$ 		& = & $\frac{2}{3}x_H + \frac{1}{3}x_\Phi$  \\[2pt] 
$d_R^i$    & {\bf 3} & {\bf 1}& $-\frac{1}{3}$ & $x_d^\prime$ 		& = & $-\frac{1}{3}x_H + \frac{1}{3}x_\Phi$  \\[2pt] 
\hline
\hline
&&&&&&\\[-12pt]
$\ell_L^i$    & {\bf 1} & {\bf 2}& $-\frac{1}{2}$ & $x_\ell^\prime$ 	& = & $- \frac{1}{2}x_H - x_\Phi$   \\[2pt] 
$e_R^i$   & {\bf 1} & {\bf 1}& $-1$   & $x_e^\prime$ 		& = & $- x_H - x_\Phi$  \\[2pt] 
\hline
\hline
$N_R^i$   & {\bf 1} & {\bf 1}& $0$   & $x_\nu^\prime$ 	& = & $- x_\Phi$ \\[2pt] 
\hline
\hline
&&&&&&\\[-12pt]
$H$         & {\bf 1} & {\bf 2}& $-\frac{1}{2}$  &  $-\frac{x_H}{2}$ 	& = & $-\frac{x_H}{2}$\hspace*{12.5mm}  \\ 
$\Phi$      & {\bf 1} & {\bf 1}& $0$  &  $2 x_\Phi$ 	& = & $2 x_\Phi$  \\ 
\hline
\hline
\end{tabular}
\end{center}
\caption{The particle content of the $U(1)_X$ scenario where i
is the family indices for the three generations.}
\label{tab1}
\end{table}
Fixing $x_\Phi=1$ and varying $x_H$, it is found that for $x_H=-2$ the $U(1)_X$ charge of the left-handed fermion doublets $(q_L^\alpha, \ell_L^\alpha)$ vanish and a $U(1)_R$ scenario \cite{Jung:2009jz,Nomura:2017tih,Nomura:2017ezy,Jana:2019mez,Seto:2020jal} 
can be realized. For $x_H=-1$ those of right-handed leptons $(e_R^\alpha)$ vanish, for $x_H=-0.5$ those of right-handed up-type quark $(u_R^\alpha)$ vanish, and finally, for $x_H=1$ those of right-handed down-type quark $(d_R^\alpha)$ vanish, respectively. 
For $x_H=0$, this scenario reproduces $U(1)_{B-L}$ scenario~\cite{Davidson:1979wr,Mohapatra:1980qe,Wetterich:1981bx,Masiero:1982fi,Buchmuller:1991ce} where left- and right-handed fermions have the same $U(1)_X$ charges. 
On the other hand, for $x_H=2$, it manifests a chiral scenario where the left- and right-handed fermions have completely different $U(1)_X$ charges.

In the general $U(1)_X$ model, the renormalizable scalar potential is given by 
\begin{align}
  V \ = \ \mu_H^2(H^\dag H) + \lambda_H^{} (H^\dag H)^2 + \mu_\Phi^2 (\Phi^\dag \Phi) + \lambda_\Phi^{} (\Phi^\dag \Phi)^2 + \lambda_{\rm mix} (H^\dag H)(\Phi^\dag \Phi)~.
\end{align}
After the general U(1)$_X$ and electroweak symmetries are broken, the scalar fields develop non-zero VEVs as
\begin{align}
  \braket{H} \ = \ \frac{1}{\sqrt{2}}\begin{pmatrix} v+h\\0 
  \end{pmatrix}~, \quad {\rm and}\quad 
  \braket{\Phi} \ =\  \frac{v_\Phi^{}+ \phi}{\sqrt{2}}~,
\end{align}
where $v=246$ GeV, and $v_\Phi^{}$ is a free parameter. In this analysis, we consider that the mixed quartic coupling $(\lambda_{\rm mix})$, that is, the mixing between the scalar fields is very small. After the breaking of general $U(1)_X$, the $U(1)_X$ gauge boson $Z^\prime$ acquires mass as  
\begin{equation}
 M_{Z^\prime}^{}=  2 g_X^{}  v_\Phi^{}~,
\end{equation}
under the limit $v_\Phi^{} \gg v$ and considering $x_\Phi=1$, and $g_X$ is the gauge coupling of the general $U(1)_X$ gauge symmetry. In this model, the neutrino mass is generated with the help of the Yukawa interactions given in Eq.~(\ref{LYk}) after the breaking of $U(1)_X$ and electroweak symmetries. We obtain the Majorana mass after the $U(1)_X$ breaking as
\bea
M_N=\frac{Y_N v_\Phi}{2\sqrt{2}}~,
\eea
and after the electroweak symmetry breaking, the Dirac mass term is generated as 
\bea
m_D=\frac{Y_\nu v}{\sqrt{2}}~,
\eea
suppressing the indices. 

\subsection{Case-II}
\begin{table}[t]
\begin{center}
\begin{tabular}{|c|c|c|c|c|c|}
\hline\hline
      &  $SU(3)_c$  & $SU(2)_L$ & $U(1)_Y$ & $U(1)_X$ \\ 
\hline
$q_{L_i}$ & {\bf 3 }    &  {\bf 2}         & $ 1/6$       &  $ x^\prime_q= (1/6) x_{H} + (1/3)$ \\
$u_{R_i}$ & {\bf 3 }    &  {\bf 1}         & $ 2/3$       & $x^\prime_u=(2/3) x_{H} + (1/3) $ \\
$d_{R_i}$ & {\bf 3 }    &  {\bf 1}         & $-1/3$       & $x^\prime_d=-(1/3) x_{H} + (1/3) $\\
\hline
\hline
$\ell_{L_i}$ & {\bf 1 }    &  {\bf 2}         & $-1/2$       & $x^\prime_\ell=(-1/2) x_{H} - 1 $ \\
$e_{R_i}$    & {\bf 1 }    &  {\bf 1}         & $-1$         & $x^\prime_e=-x_{H} - 1 $ \\
\hline
\hline
$N_{R_{1,2}}$    & {\bf 1 }    &  {\bf 1}         &$0$                    & $x^\prime_\nu=- 4 $ \\ 
$N_{R_3}$    & {\bf 1 }    &  {\bf 1}         &$0$                           & $x^{\prime\prime}_{\nu}=+ 5 $   \\
\hline
\hline
$H_1$            & {\bf 1 }    &  {\bf 2}         & $- 1/2$       & $(-1/2) x_{H}$ \\  
$H_2$            & {\bf 1 }       &  {\bf 2}       &$ -1/2$                  & $(-1/2) x_{H}+3 $  \\ 
$\Phi_1$            & {\bf 1 }       &  {\bf 1}       &$ 0$                  & $ +8  $  \\ 
$\Phi_2$            & {\bf 1 }       &  {\bf 1}       &$ 0$                  & $ -10 $  \\ 
$\Phi_3$          & {\bf 1 }       &  {\bf 1}       &$ 0$                  & $ -3 $  \\
\hline\hline
\end{tabular}
\end{center}
\caption{
Particle content of the `alternative' general $U(1)$ extension of the SM where $i$ denotes the generation index. 
}
\label{tab2}
\end{table}   

There is another interesting $U(1)_X$ extension of the SM whose minimal particle content is shown in Tab.~\ref{tab2}. We call it an alternative U$(1)_X$ scenario.  The $U(1)_X$ charge $x_H$ is a real parameter and the $U(1)_X$ coupling $g_X$ is a free parameter. The RHNs in this model are differently charged under the $U(1)_X$.
In this model first two generations of RHNs have charge $-4$ whereas the third one has a charge $+5$. This non-universal charge assignment is a unique choice in order to cancel all the anomalies \cite{Montero:2007cd}. The $U(1)_X$ charges of the SM fermions are the same for three generations. The general charges can be related to each other from the following gauge and mixed gauge-gravity anomaly cancellation conditions  
\begin{align}
{\rm U}(1)_X \otimes \left[ {\rm SU}(3)_C \right]^2&\ :&
			2x_q^\prime - x_u^\prime - x_d^\prime &\ =\  0~, \nonumber \\
{\rm U}(1)_X \otimes \left[ {\rm SU}(2)_L \right]^2&\ :&
			3x_q^\prime + x_\ell^\prime &\ =\  0~, \nonumber \\
{\rm U}(1)_X \otimes \left[ {\rm U}(1)_Y \right]^2&\ :&
			x_q^\prime - 8x_u^\prime - 2x_d^\prime + 3x_\ell^\prime - 6x_e^\prime &\ =\  0~, \nonumber \\
\left[ {\rm U}(1)_X \right]^2 \otimes {\rm U}(1)_Y &\ :&
			{x_q^\prime}^2 - {2x_u^\prime}^2 + {x_d^\prime}^2 - {x_\ell^\prime}^2 + {x_e^\prime}^2 &\ =\  0~, \nonumber \\
\left[ {\rm U}(1)_X \right]^3&\ :&
			3({6x_q^\prime}^3 - {3x_u^\prime}^3 - {3x_d^\prime}^3 + {2x_\ell^\prime}^3-{x_e^\prime}^3) - 2 x_\nu^{\prime^3}-x_\nu^{\prime \prime^3}   &\ =\  0~, \nonumber \\
{\rm U}(1)_X \otimes \left[ {\rm grav.} \right]^2&\ :&
			3(6x_q^\prime - 3x_u^\prime - 3x_d^\prime + 2x_\ell^\prime-x_e^\prime)-2 x_\nu^{\prime}-x_\nu^{\prime \prime}  &\ =\  0~, 
\label{anom-f}
\end{align}
respectively. Using SM $\otimes~U(1)_X$ gauge symmetry we write the Yukawa interactions as 
\begin{eqnarray}
- L^{\text{lepton}}_Y &=& \bar{\ell}_L y_l \tilde{H}_1 e_R + \sum_{i=1}^3 \sum_{j=1}^2 Y_D^{ij} \bar{\ell}_{Li} H_2 N_{R_j} + \frac{1}{2}\sum_{k=1}^2 Y_{2}^{k} \bar{N}^C_{R_{k}}\Phi_1 N_{R_k} + \frac{1}{2} Y_3 \bar{N}^C_{R_3}\Phi_2 N_{R_3} + \text{H.c.} \nonumber \\
-L^{\text{quark}}_Y &=& \bar{Q}_L y_d \tilde{H}_1d_R + \bar{Q}_L y_u H_1u_R  + \text{H.c.}
\label{LYukawa}
\end{eqnarray}

In this model we introduce two Higgs doublets $(H_1, H_2)$ and three additional SM-singlet scalars $(\Phi_{1,2,3})$. The Higgs doublet $H_2$ is responsible for the generation of the Dirac mass term for $N_{R_{1,2}}$. The SM-singlet scalar $\Phi_{1}$ is responsible for the generation of the Majorana mass term of $N_{R_{1,2}}$ after the $U(1)_X$ breaking. The Majorana mass term of $N_{R_{3}}$ is generated from the VEV of $\Phi_2$, however, there is no Dirac mass term for $N_{R_{3}}$ due to the preservation of $U(1)_X$ symmetry. Hence $N_{R_3}$ does not participate in the neutrino mass generation mechanism. The relevant part of the interaction Lagrangian of the RHNs is given by
\bea
-\mathcal{L} _{\rm int}& \ \supset \ & \sum_{i=1}^{3} \sum_{j=1}^{2} Y_{1}^{ij} \overline{\ell_{L_i}} H_2 N_{R_j}+\frac{1}{2} \sum_{k=1}^{2} Y_{2}^{k}  \overline{N_{R_k}^{C}}\Phi_1 N_{R_k} 
+\frac{1}{2} Y_{3} \overline{N_{R_3}^{C}} \Phi_2  N_{R_3}+ \rm{H. c.} \, ,
\label{ExoticYukawa}
\eea 
where we have assumed a basis in which $Y_2$ is diagonal, without the loss of generality. The scalar potential is given by 
\bea
  V&\ =\ &
m_{H_1}^2 (H_1^\dagger H_1) + \lambda_{H_1}  (H_1^\dagger H_1)^2 + m_{H_2}^2 (H_2^\dagger H_2) + \lambda_{H_2}  (H_2^\dagger H_2)^2 \nonumber \\
&& + m_{\Phi_1}^2 (\Phi_1^\dagger \Phi_1) + \lambda_1  (\Phi_1^\dagger \Phi_1)^2 
+ m_{\Phi_2}^2 (\Phi_2^\dagger \Phi_2) + \lambda_2   (\Phi_2^\dagger \Phi_2)^2 \nonumber \\
&&+ m_{\Phi_3}^2 (\Phi_3^\dagger \Phi_3) + \lambda_3   (\Phi_3^\dagger \Phi_3)^2 
+ ( \mu \Phi_3 (H_1^\dagger H_2) + {\rm H.c.} )  \nonumber \\
&&+ \lambda_4 (H_1^\dagger H_1) (H_2^\dagger H_2)+ \lambda_5 (H_1^\dagger H_2) (H_2^\dagger H_1) +\lambda_6 (H_1^\dagger H_1) (\Phi_1^\dagger \Phi_1)\nonumber \\
&&+ \lambda_7 (H_1^\dagger H_1) (\Phi_2^\dagger \Phi_2)+ \lambda_8 (H_1^\dagger H_2) (\Phi_3^\dagger \Phi_3) +\lambda_9 (H_2^\dagger H_2) (\Phi_1^\dagger \Phi_1)  \nonumber \\
&&+ \lambda_{10} (H_1^\dagger H_1) (\Phi_2^\dagger \Phi_2)+ \lambda_{11} (H_1^\dagger H_2) (\Phi_3^\dagger \Phi_3)+  \lambda_{12} (\Phi_1^\dagger \Phi_1) (\Phi_2^\dagger \Phi_2) \nonumber \\
&&+ \lambda_{13} (\Phi_2^\dagger \Phi_2) (\Phi_3^\dagger \Phi_3)+ \lambda_{14} (\Phi_3^\dagger \Phi_3) (\Phi_1^\dagger \Phi_1).
\label{HiggsPotential-2}
\eea
We choose suitable parameters for the Higgs fields to develop their respective VEVs: 
\bea
  \langle H_1 \rangle \ = \  \frac{1}{\sqrt 2}\left(  \begin{array}{c}  
    v_{h_1} \\
    0 \end{array}
\right),   \; 
\langle H_2 \rangle \ = \   \frac{1}{\sqrt{2}} \left(  \begin{array}{c}  
    v_{h_2}\\
    0 \end{array}
\right),  
\langle \Phi_1 \rangle \ = \  \frac{v_{1}}{\sqrt{2}},  \; 
\langle \Phi_2 \rangle \ = \  \frac{v_{2}}{\sqrt{2}},  \; 
\langle \Phi_3 \rangle \ = \  \frac{v_{3}}{\sqrt{2}},~~~~ 
\eea   
with the condition, $v_{h_1}^2 + v_{h_2}^2 = (246 \,  {\rm GeV})^2$. 
We consider negligibly small mixed-quartic couplings between the Higgs doublets 
  and the SM singlets for simplicity, so that the Higgs singlet sector is effectively separated from
  the Higgs doublets. It ensures that any higher-order mixing effect between the three generations of the RHNs after $U(1)_X$ symmetry-breaking will be extremely suppressed.
  The singlet and doublet Higgs sectors communicate only through the triple coupling $\Phi_3 (H_1^\dagger H_2)+{\rm H.c.}$ 
Taking the collider constraints into account $v_1^2 + v_2^2+ v_3^2 \gg v_{h_1}^2 + v_{h_2}^2$, 
  the triple coupling has no significant effect on determining the VEVs $(v_{1, 2, 3})$ of the SM-singlet scalars $(\Phi_1, \Phi_2, \Phi_3)$
  when we arrange the parameters in the scalar potential to have the VEVs of the SM-singlet scalars almost same $(v_1 \sim v_2 \sim v_3)$ 
  and $\mu < v_1$. The third SM-singlet scalar, $\Phi_3$,  can be used as a spurion for the Higgs doublet sector  
  which generates the mixing between $H_1$ and $H_2$ through the term $\mu \Phi_3 (H_1^\dagger H_2)+{\rm H.c.}$ Using $\langle \Phi_3 \rangle =\frac{v_3}{\sqrt{2}}$ the mixing mass term becomes $m_{\rm{mix}}^2=\frac{\mu v_3}{\sqrt{2}}$.
  Hence the Higgs doublet sector potential effectively becomes the Higgs potential of the two Higgs doublet model.  
There is no mixing mass term among $\Phi_{1,2,3}$ due to the $U(1)_X$ symmetry. As a result there are two physical Nambu-Goldstone (NG) modes present in our model. 
These NG modes are originated from the SM singlet scalars and they are not phenomenologically dangerous. 
We consider the SM-singlet scalars heavier than $Z^\prime$, so that $Z^\prime$ cannot decay into the NG modes. 
After the $U(1)_X$ symmetry is broken the $Z^\prime$ boson acquires the mass term as
\bea
 M_{Z^\prime} = g_X \sqrt{64 v_{1}^2+ 100 v_{2}^2+ 9v_3^2 +\frac{1}{4} x_H^2 v_{h_{1}}^2 + \left(-\frac{1}{2} x_H +3\right)^2  v_{h_{2}}^2}
\simeq g_X \sqrt{64 v_{1}^2+ 100 v_{2}^2+ 9 v_{3}^2}~~~~~~~.
\label{masses-Alt}   
\eea 
and the Majorana masses of the RHNs are generated as
\bea
 M_{N_{1,2}}=\frac{Y_2^{1,2}}{\sqrt{2}} v_1,\;\; M_{N_3} = \frac{Y_3^{3}}{\sqrt{2}} v_2,
\label{mN3II} 
\eea
using the collider constraints to set $(v_1^2 + v_2^2+ v_3^2) \gg (v_{h_{1}}^2 + v_{h_{2}}^2)$.  
The Dirac mass terms of the neutrinos are generated by $\langle H_2 \rangle$: 
\bea
M_{D}^{ij} \ = \ \frac{Y_{1}^{ij}}{\sqrt{2}} \, v_{h_{2}} \, , \label{mDII}
\eea
after which the seesaw mechanism is implemented.
Because of the $U(1)_X$ charges, only two RHNs~($N_{R_{1,2}}$) are involved in the minimal seesaw mechanism 
  \cite{Smirnov:1993af,King:1999mb,Frampton:2002qc,Ibarra:2003up}    
   while the third RHN ($N_{R_3}$) has no direct interaction with the SM sector. 
   Hence it can be a potential Dark Matter (DM) candidate.
  Due to the $U(1)_X$ symmetry the Higgs doublet $H_1$ has no coupling with the RHNs and
the neutrino Dirac masses are generated by the VEV of $H_2$ as mentioned in Eq.~(\ref{mDII}). This structure can be considered as a type of the 
neutrinophilic two Higgs Doublet Model (2HDM) \cite{Ma:2000cc,Wang:2006jy,Gabriel:2006ns,Davidson:2009ha,Haba:2010zi}. 
In Eq.~(\ref{HiggsPotential-2}) we may consider $0 < m_{\rm mix}^2 = \frac{\mu v_3}{\sqrt{2}} \ll m_{\Phi_3}^2$  
  which leads to $v_{h_{2}} \sim m_{\rm mix}^2 v_{h_{1}}/m_{\Phi_{3}}^2 \ll v_{h_{1}}$ \cite{Ma:2000cc}. Variations of $x_H$ will be same as the $U(1)_X$ case with $x_\Phi=1$.  We refer this case as `alternative' scenario henceforth. 
\subsection{Neutrino mass}
In these two cases, Majorana mass term of the RHNs are generated after the breaking of $U(1)_X$ symmetry and Dirac mass term of the RHNs are generated after the breaking of electroweak symmetry. These masses induce well-known seesaw mechanism to generate the light neutrino mass and flavor mixings. The mass matrix for the light and heavy neutrinos is given as 
\begin{equation}
\mathcal{M}= \begin{pmatrix} 0&m_D^{}\\m_D^T&M_N^{} \end{pmatrix}~.
\label{num-1}
\end{equation}
By block-diagonalizing the above neutrino mass matrix, the mass matrix for the light neutrinos is obtained as
\begin{equation}
m_\nu \simeq - m_D^{} M_N^{-1} m_D^T~,
\end{equation}
through the seesaw mechanism.
\subsection{$Z^\prime$ interactions with the fermions}
In this model, the $U(1)_X$ gauge interactions of the left- and right-handed fermions depend on their general $U(1)_X$ charges, and hence, manifest chiral nature. 
The interactions between the fermions with the $Z^\prime$ are given as follows~\footnote{Note that $Z'$ can also decays to channels such as $h_{1,2}Z$ and $W^+W^-$ but are kinematically forbidden for our interested mass range of $Z'$.}:
\bea
\mathcal{L}_{\rm int} = -g_X (\overline{f}\gamma^\mu q_{f_{L}^{}}^{} P_L^{} f+ \overline{f}\gamma^\mu q_{f_{R}^{}}^{}  P_R^{} f) Z_\mu^\prime~,
\label{Lag1}
\eea
where $P_{L(R)}^{}= (1 \pm \gamma_5)/2$ is the left- (right-)handed projections. 
In Eq.~\eqref{Lag1}, $q_{f_{L(R)}^{}}^{}$ is the corresponding general U(1)$_X$ charge of the left- (right-)handed fermion $(f_{L(R)})$ and is expressed by $x_H$, as shown in Tab.~\ref{tab1}. The partial decay widths of $Z^\prime$ into a pair of charged fermions are calculated as 
\begin{align}
\label{eq:width-ll}
    \Gamma(Z' \to \bar{f} f)
    &= N_C^{} \frac{M_{Z^\prime}^{} g_{X}^2}{24 \pi} \left[ \left( q_{f_L^{}}^2 + q_{f_R^{}}^2 \right) \left( 1 - \frac{m_f^2}{M_{Z^\prime}^2} \right) + 6 q_{f_L^{}}^{} q_{f_R^{}}^{} \frac{m_f^2}{M_{Z^\prime}^2} \right] \left( 1 - 4 \frac{m_f^2}{M_{Z^\prime}^2} \right)^{\frac{1}{2}}~,
    \end{align}    
where $m_f$ stands for the mass of the SM fermions, $N_C^{}$ is the color factor which is $1 (3)$ for the SM leptons (quarks). That into a pair of light neutrinos $(\nu)$ is calculated as
\begin{align}   
\label{eq:width-nunu}
    \Gamma(Z' \to \nu \nu)
    = \frac{M_{Z^\prime}^{} g_{X}^2}{24 \pi} q_{\nu_L^{}}^2~,
\end{align} 
neglecting the tiny mass of the light neutrinos. In this model, the $Z^\prime$ gauge boson can also decay into a pair of heavy Majorana neutrinos following
\bea
\mathcal{L}_N= -\frac{1}{2}g_X q_{N_R} \overline{N} \gamma_\mu \gamma_5 N Z_{\mu}^\prime.
\label{neut}
\eea
where $q_{N_R}$ is the $U(1)_X$ charge of the RHNs. When $Z^\prime$ is heavier than twice the mass of heavy neutrinos $(M_N)$, it can decay into a pair of RHNs. The corresponding partial decay width is given by
\begin{align}
\label{eq:width-NN}
    \Gamma(Z^\prime \to N N)
    = \frac{M_{Z^\prime}^{} g_{X}^2}{24 \pi} q_{N_R^{}}^2 \left( 1 - 4 \frac{M_N^2}{M_{Z^\prime}^2} \right)^{\frac{3}{2}}~,
\end{align}
with $q_{N_R^{}}^{}$ is the general $U(1)_X$ charge of the heavy neutrinos, and in our case, it is $q_{N_R^{}}^{}=-x_\Phi=-1$. If we consider $M_N > M_{Z^\prime}/2$, then $Z^\prime \to N N$ mode will be kinematically forbidden. In this paper, we will study the effect from the presence of the RHNs and the decay of the RHNs after they are pair-produced from $Z^\prime$. 

\begin{figure}[h]
\includegraphics[scale=0.52]{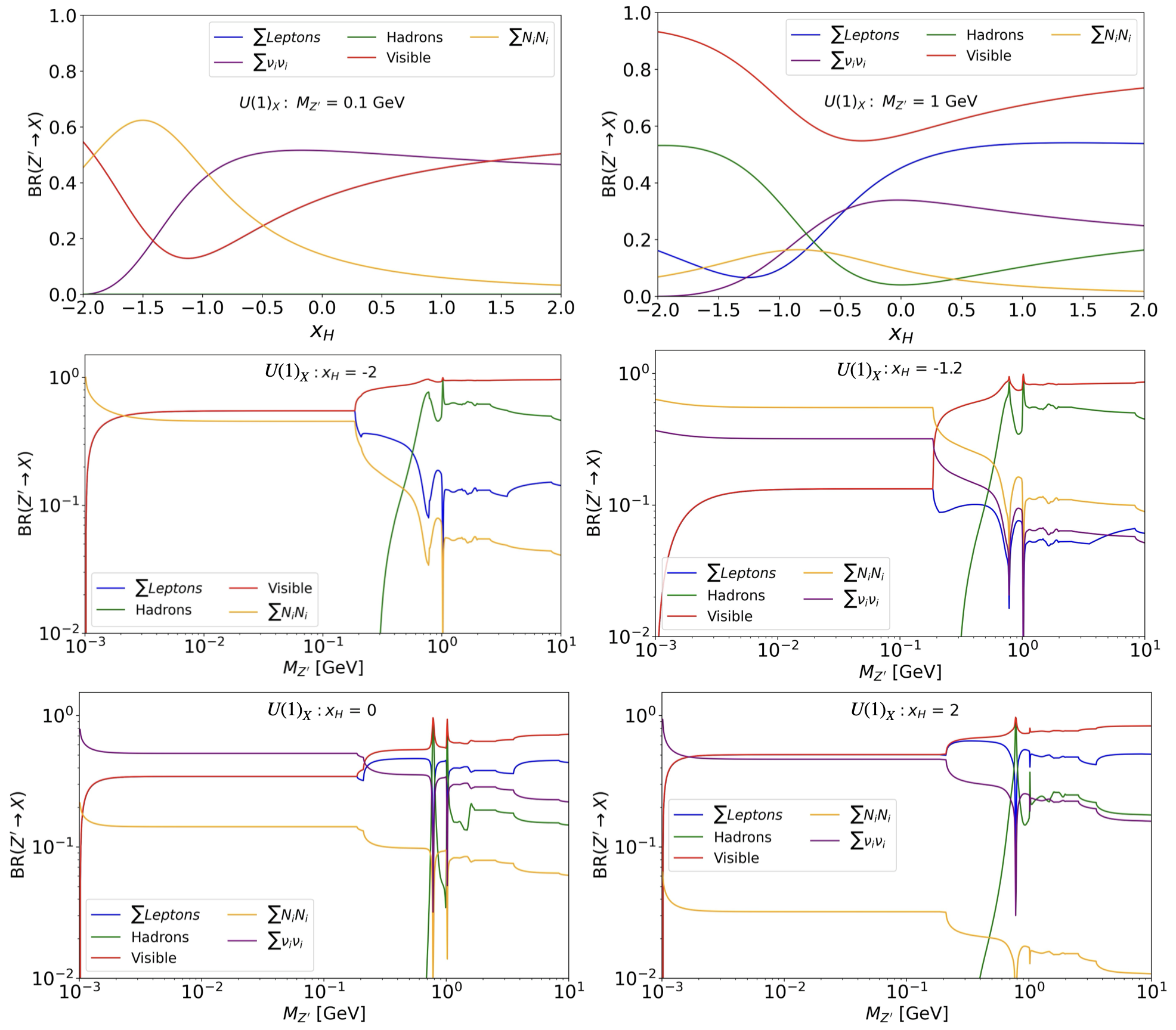}
\caption{Branching ratios of $Z^\prime$ into different modes considering $x_\Phi=1$ for U$(1)_X$ case. In the top row branching ratios are shown as a function of $x_H$ for $M_{Z^\prime}=0.1$ (1) GeV in the left (right) panel considering $M_{N_{1,2}} = M_{Z^\prime}/3$ and $M_{N_{3}} > M_{Z^\prime}/2$. In the middle and bottom rows, branching ratios are shown as a function of $M_{Z^\prime}$ for fixed $x_H$.}
\label{branching-Zp}
\end{figure}

In this analysis we consider a scenario for the two cases where two generations of the RHNs participate in the neutrino mass generation mechanism and flavor mixing where the third one could be considered as a potential DM candidate. In Fig.~\ref{branching-Zp} we show the branching ratio of $Z^\prime$ into different modes for $U(1)_X$ considering $M_{N_{1,2}}= M_{Z^\prime}/3$ and $M_{N_{3}} > M_{Z^\prime}/2$. In the top panel we show branching ratios of $Z^{\prime}$ into different modes for different $x_H$ fixing $M_{Z^\prime}=0.1$(1) GeV in the top left (right) panel. In the $U(1)_X$ scenario $Z^\prime \to NN$ modes becomes dominant when $x_H=-1.2$ for $M_{Z^\prime}=0.1$ GeV. We show the respective branching ratios of $Z^\prime$ as a function of $M_{Z^\prime}$ for different $x_H$ in the middle and bottom panels of Fig.~\ref{branching-Zp} considering $x_H=-2$, $-1.2$, $0$ and $2$ respectively. We find that in case of $x_H=-2$, BR$(Z^\prime \to N N)$ dominates over the other modes when $M_{Z^\prime} \leq 0.002$ GeV. BR$(Z^\prime \to N N)$ becomes almost $50\%$ for $0.002$ GeV $\leq M_{Z^\prime} \leq 0.02$ GeV. For $M_{Z^\prime} \geq 0.02$ GeV, $Z^\prime$ starts decaying into hadronic modes resulting decrements in the RHN and other leptonic modes. For $M_{Z^\prime} \geq 1$ GeV, BR$(Z^\prime)\to NN$ varies between $4\%-6\%$ up to $M_{Z^\prime}=10$ GeV.
\begin{figure}[h]
\includegraphics[scale=0.52]{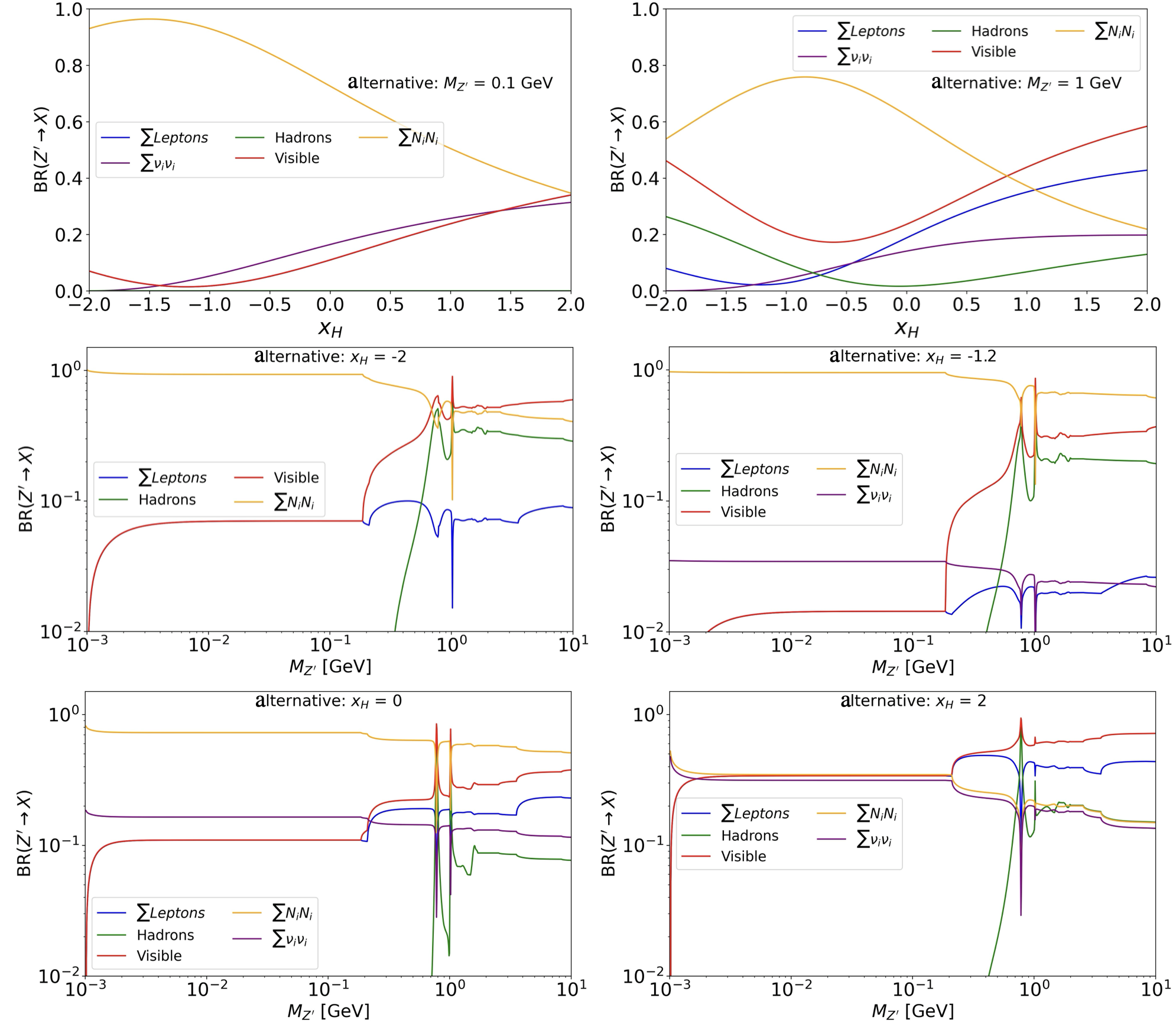}
\caption{Branching ratios of $Z^\prime$ into different modes for the `alternative' case. In the top row branching ratios are shown as a function of $x_H$ for $M_{Z^\prime}=0.1$ (1) GeV in the left (right) panel considering $M_{N_{1,2}} = M_{Z^\prime}/3$ and $M_{N_{3}} > M_{Z^\prime}/2$. In the middle and bottom rows, branching ratios are shown as a function of $M_{Z^\prime}$ for fixed $x_H$.}
\label{branching-Zp-1}
\end{figure}
In case of $x_H=-1.2$, BR$(Z^\prime \to N N)\simeq 60\%$ for $0.001$ GeV $\leq M_{Z^\prime} \leq 0.2$ GeV dominates over the other modes whereas it drops below 20$\%$ up to $M_{Z^\prime}=1$ GeV due to the hadronic decays of $Z^\prime$. However, BR$(Z^\prime \to N N)\simeq 10\%$ up to $M_{Z^\prime}=10$ GeV being subdominat to the hadronic mode and dominant over the leptonic modes. In case of $x_H=0$, BR$(Z^\prime \to N N)$ varies between $8\%-20\%$ for $0.001$ GeV $\leq M_{Z^\prime} \leq 10$ GeV being subdominant under leptonic and hadronic modes. Same scenario could be observed for $x_H=2$ where BR$(Z^\prime \to N N)$ varies between $6\%-1\%$ for $0.001$ GeV $\leq M_{Z^\prime} \leq 10$ GeV where other decay modes from $Z^\prime$ dominate over the RHNs.

In Fig.~\ref{branching-Zp-1} we consider the `alternative' case where $U(1)_X$ charge of each of the first two generations of the RHNs is $-4$ and the masses of these RHNs are $M_{N_{1,2}}=M_{Z^\prime}/3$. We show the branching ratio of $Z^\prime$ into different modes as a function of $x_H$ for $M_{Z^\prime}=0.1$(1) GeV in the top left (right) panel. We find that BR$(Z^\prime \to N N)$ varies between $40\%-100\%$ for $-2\leq x_H\leq 2$ when $M_{Z^\prime}=0.1$ GeV and that for $M_{Z^\prime}=1$ GeV varies between roughly $37\%-75\%$ for $-2\leq x_H\leq 1.2$. In the middle and lower panels we show branching ratios of $Z^\prime$ into different modes as a function of $M_{Z^\prime}$ for $x_H=-2,$ $-1.2$, $0$ and $2$. We find that for $x_H=-2$, $-1.2$ and $0$, BR$(Z^\prime \to NN)$ dominates and  varies between $70\%-100\%$ in mass window of $0.7$ GeV $\leq M_{Z^\prime} \leq 1$ GeV.  However, the branching ratio of $Z^\prime \to NN$ varies between $15\%-50\%$ for $x_H=2$ for $0.001$ GeV $\leq M_{Z^\prime} \leq 10$ GeV apart from the peaks near the hadronic mass. Important to mention that the visible decay of $Z^\prime$ comprises of all possible decay modes including charged leptons, hadrons, charm, strange, bottom quarks except neutrinos as shown by the red solid line in Figs.~\ref{branching-Zp} and \ref{branching-Zp-1} respectively. To calculate branching ratios of $Z^\prime$ we implemented our models in   DarkCast \cite{Baruch:2022esd}.  

\begin{figure}[h]
\includegraphics[scale=0.33]{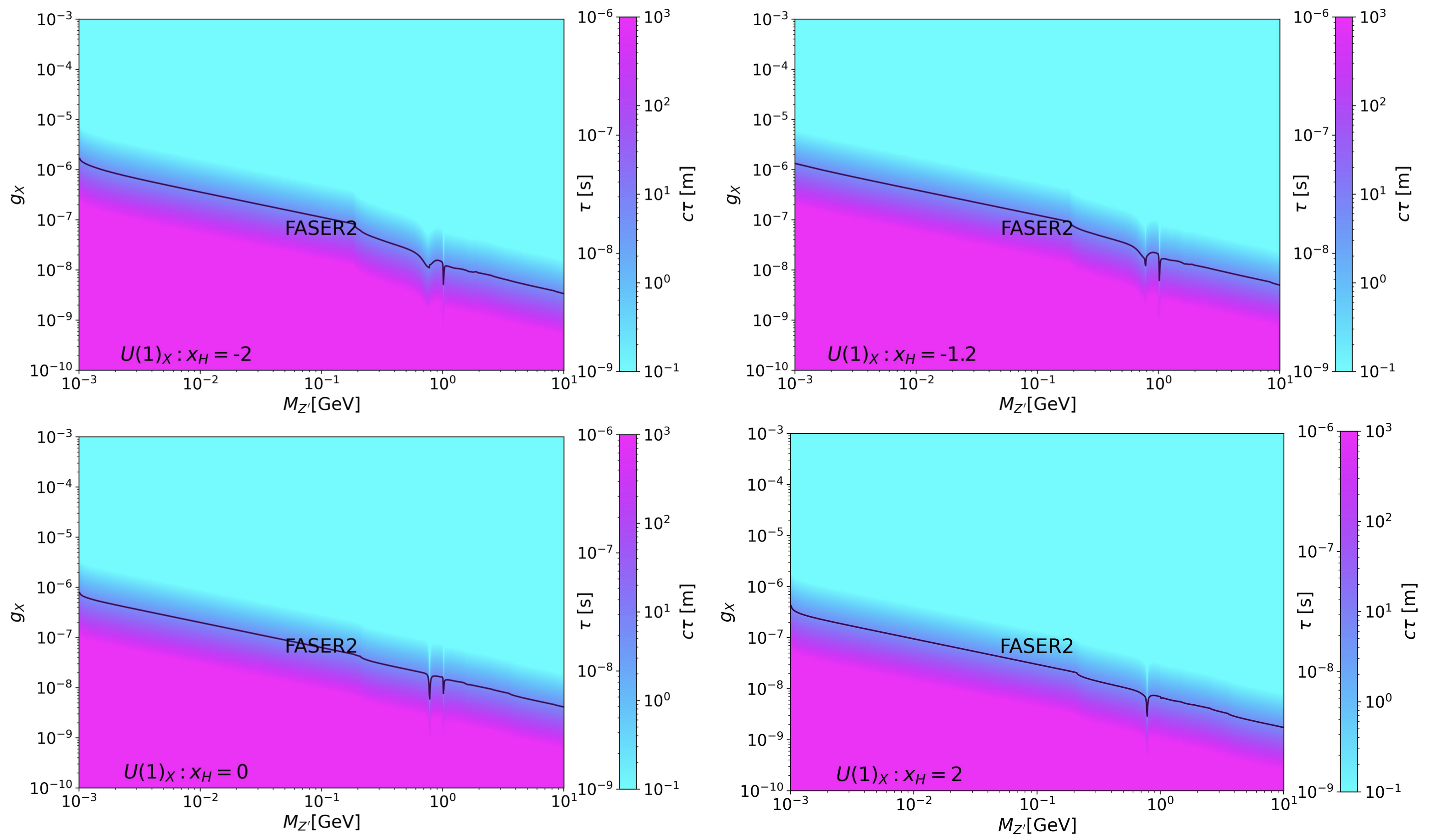}
\caption{Decay length of $Z'$ as a function of gauge coupling $(g_X)$, $M_{Z'}$ and $x_H$ for $U(1)_X$ scenario considering the decay of $Z^\prime$ at rest frame. The black contour stands for the decay length $L_{Z'}=620$ m relevant for FASER2 detector.}
\label{fig:decaylength-Zp}
\end{figure}
\begin{figure}[h]
\centering
\includegraphics[scale=0.35]{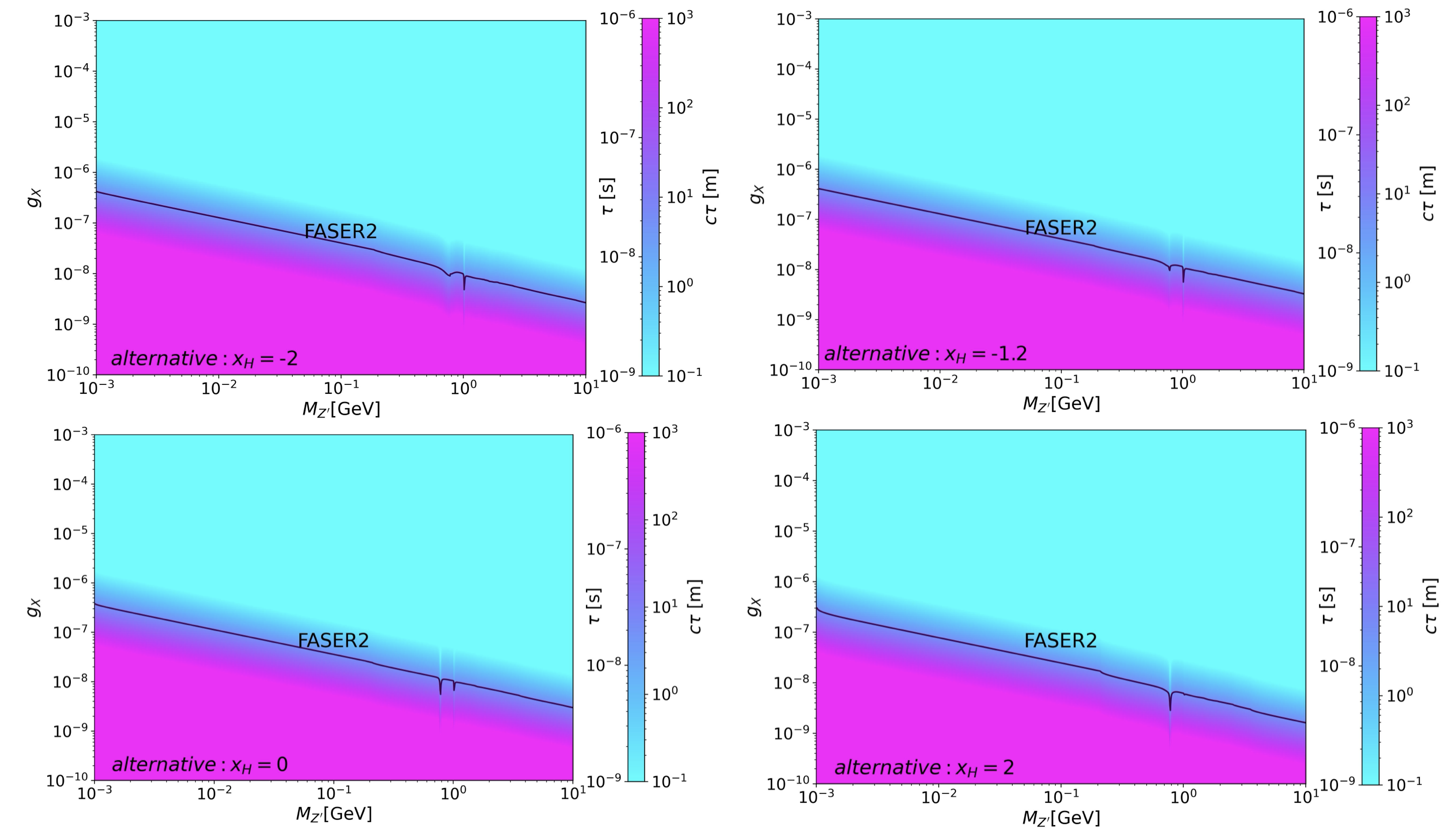}
\caption{Decay length of $Z'$ as a function of gauge coupling $(g_X)$, $M_{Z'}$ and $x_H$ for the `alternative' scenario considering the decay of $Z^\prime$ at rest frame. The black contour stands for the decay length $L_{Z'}=620$ m relevant for FASER2 detector.}
\label{fig:decaylength-Zp-1}
\end{figure}

Summing over the partial decay widths given in Eqs.~(\ref{eq:width-ll}), (\ref{eq:width-nunu}), and (\ref{eq:width-NN}), the total decay width of $Z^\prime$ is obtained as a function of $M_{Z^\prime}$, $x_H$ and $g_X$ which is finally used to calculate the decay length of $Z^\prime$ at rest frame as 
\bea
c\tau=\frac{1.97\times 10^{-16}\,{\rm GeV}}{\Gamma[M_{Z^\prime}, g_X, x_H]} \,{\rm [m]}~.
\label{DL}
\eea
In Fig.~\ref{fig:decaylength-Zp}, we show the rest frame decay length of $Z'$ as functions of gauge coupling $g_X$, $M_{Z'}$ and $x_H$ with corresponding lifetime for the $U(1)_X$ case. The solid black line shown by FASER2 is the decay length of $Z^\prime$ at 620 m. The decay length of $Z^\prime$ strongly depend on $g_X$. From magenta to cyan the $Z^\prime$ becomes more short-lived. In Fig.~\ref{fig:decaylength-Zp-1} we show the theoretical decay length of $Z^\prime$ in the `alternative' scenario where total decay width~(decay length) of $Z^\prime$ increases~(decreases) due to the corresponding $U(1)_X$ charges of the RHNs. Therefore to obtain the decay length of $Z^\prime$ about 620 m at FASER2, the $U(1)_X$ gauge coupling $g_X$ will decreases.
\subsection{RHN interactions and partial decay modes}
The RHNs (HNLs) are singlet under the SM gauge group, RHNs do not interact with the SM sector directly. 
However, from the seesaw mechanism, we can write the light neutrino flavor eigenstate $(\nu)$ in terms of the mass eigenstates of the light $(\nu_m)$ and heavy $(N_m)$ neutrinos as 
\bea 
  \nu \simeq  \nu_m  + V_{\ell N} N_m~,  
\eea 
where $V_{\ell N}(=m_D/m_N)$ is the mixing between the light and heavy mass eigenstates, and we assume $|V_{\ell N}| \ll 1$. 
Hence, we write the charged current (CC) and neutral current (NC) interactions in terms of the neutrino mass eigenstates. 
The CC interaction is given by 
\bea 
\mathcal{L}_{\rm CC} \supset 
 -\frac{g}{\sqrt{2}} W_{\mu}
  \bar{\ell} \gamma^{\mu} P_L   V_{\ell N} N_m  + {\rm H.c.}, 
\label{CC}
\eea
where $\ell$ denotes the three generations of the charged leptons in the vector form, and 
$P_L =\frac{1}{2} (1- \gamma_5)$ is the projection operator. Similarly, the NC interaction is given by 
\bea 
\mathcal{L}_{\rm NC} \supset 
 -\frac{g}{2 \cos\theta_{\rm W}}  Z_{\mu} 
\left[ 
  \overline{N_m} \gamma^{\mu} P_L  |V_{\ell N}|^2 N_m 
+ \left\{ 
  \overline{\nu_m} \gamma^{\mu} P_L V_{\ell N}  N_m 
  + {\rm H.c.} \right\} 
\right] , 
\label{NC}
\eea
where $\theta_{\rm W}$ is the Weinberg mixing angle and $\sin^2\theta_W = 0.2229$. 
From Eqs.~\eqref{CC} and \eqref{NC}, the RHNs can decay into the SM particles through the mixing. 
The RHNs decay into purely leptonic modes as $N \to \nu \ell^+ \ell^-$, $N \to 3\nu$ with charged $(\ell^{\pm})$ and neutral $(\nu)$ leptons and semileptonic modes as $N \to \nu \mathscr{H}^{0}$, $N \to \ell^\pm \mathscr{H}^{\mp}$ with neutral (charged) meson ($\mathscr{H}^{0(\pm)}$), respectively. 
The corresponding partial decay widths of all possible modes are calculated in \cite{Bloom:1970xb,Bloom:1971ye,Braaten:1991qm,Cvetic:2001sn,Dib:2000wm,Ivanov:2004ch,Gribanov:2001vv,Cvetic:2010rw,Helo:2010cw}.
For the case that a virtual $W$ boson decays into leptons, the partial decay width is given by
\bea
\Gamma(N\to  \ell_1^- \ell_2^+ \nu_{\ell_2})=|V_{\ell_1 N}|^2\frac{G_F^2}{16\pi^3}M_N^5(1-\delta_{\ell_1 \ell_2})I_1(y_{\nu_{\ell_2}},y_{\ell_1}, y_{\ell_2})~, 
\label{nul1l2}
\eea
with $G_F$ being the Fermi constant, $\delta_{\ell_1 \ell_2}$ the Kronecker delta, and $y_i \equiv m_i / M_N$.
For the case that a virtual $Z$ boson decays into a charged lepton pair, the partial decay width is given by
\begin{align}
\Gamma(N\to \nu_{\ell_1} \ell_{2}^{-} \ell_{2}^{+}) =&
|V_{\ell_1 N}|^2\frac{G_F^2}{8\pi^3}M_N^5 \Big[2(g_L^\ell g_R^\ell+g_R^\ell\delta_{\ell_1 \ell_2})I_2(y_{\nu_{\ell_1}},y_{\ell_2}, y_{\ell_2}) \nonumber \\
&\hspace{30mm} + \left( (g_L^\ell)^2+(g_R^\ell)^2+(1+2g_L^\ell)\delta_{\ell_1 \ell_2} \right)
         I_1(y_{\nu_{\ell_1}},y_{\ell_2}, y_{\ell_2})\Big]~, 
\label{nuiljlj}
\end{align}
with $g_L^\ell = - 1/2 + \sin^2\theta_{\rm W}$ and $g_R^\ell = \sin^2\theta_{\rm W}$. Note that in the above we also add the process $N\to \bar{\nu}_{\ell_1}\ell_2^-\ell_2^+$.
Similarly, the RHN decays to three light neutrinos through a virtual $Z$ boson, and the corresponding partial decay width is given by
\bea
    \sum_{\ell_2=e,\mu,\tau}^{} \Gamma(N\to \nu_{\ell_1}\nu_{\ell_2}\bar{\nu}_{\ell_2})=|V_{\ell_1 N}|^2\frac{G_F^2}{96\pi^3}M_N^5~.
    \label{3nu}
\eea
Again in the above we add the process $N\to \bar{\nu}_{\ell_1}\nu_{\ell_2}\bar{\nu}_{\ell_2}$. In Eq.~\eqref{3nu}, we ignore the neutrino mass due to smallness. The partial decay widths of RHN into charged and neutral pseudoscalar mesons $(P^{+,0})$ is given as
\begin{align}
    \Gamma(N\to \ell^-_{1} P^+) &= |V_{\ell_1 N}|^2\frac{G_F^2}{16\pi}M_N^3f_P^2|V_P|^2F_P(y_{\ell}, y_P)~,
    \label{l1p} \\
    \Gamma(N\to \nu_{\ell_{1}} P^0) &= |V_{\ell_1 N}|^2\frac{G_F^2}{2\pi}M_N^3f_P^2 \kappa_P^2 F_P(y_{\nu_\ell},y_P)~,
    \label{nup}
\end{align}
where $f_P$ is the decay constant of the pseudoscalar $P$, and $V_P$ is the corresponding elements of the Cabibbo–Kobayashi –Maskawa (CKM) matrix as listed in Tab.~\ref{tab:mesdec} for each meson. $\kappa_P$ is the corresponding NC coupling of pseudoscalar meson, defined
as
\begin{align}
\kappa_{\pi^{0}}=-\frac{1}{2\sqrt{2}},\,\,\kappa_{\eta}=-\frac{1}{2\sqrt{6}}\,\,\text{and  } \kappa_{K^0}=\frac{1}{4}.
\end{align}
\begin{figure}[htb!]
\centering
\includegraphics[scale=0.4]{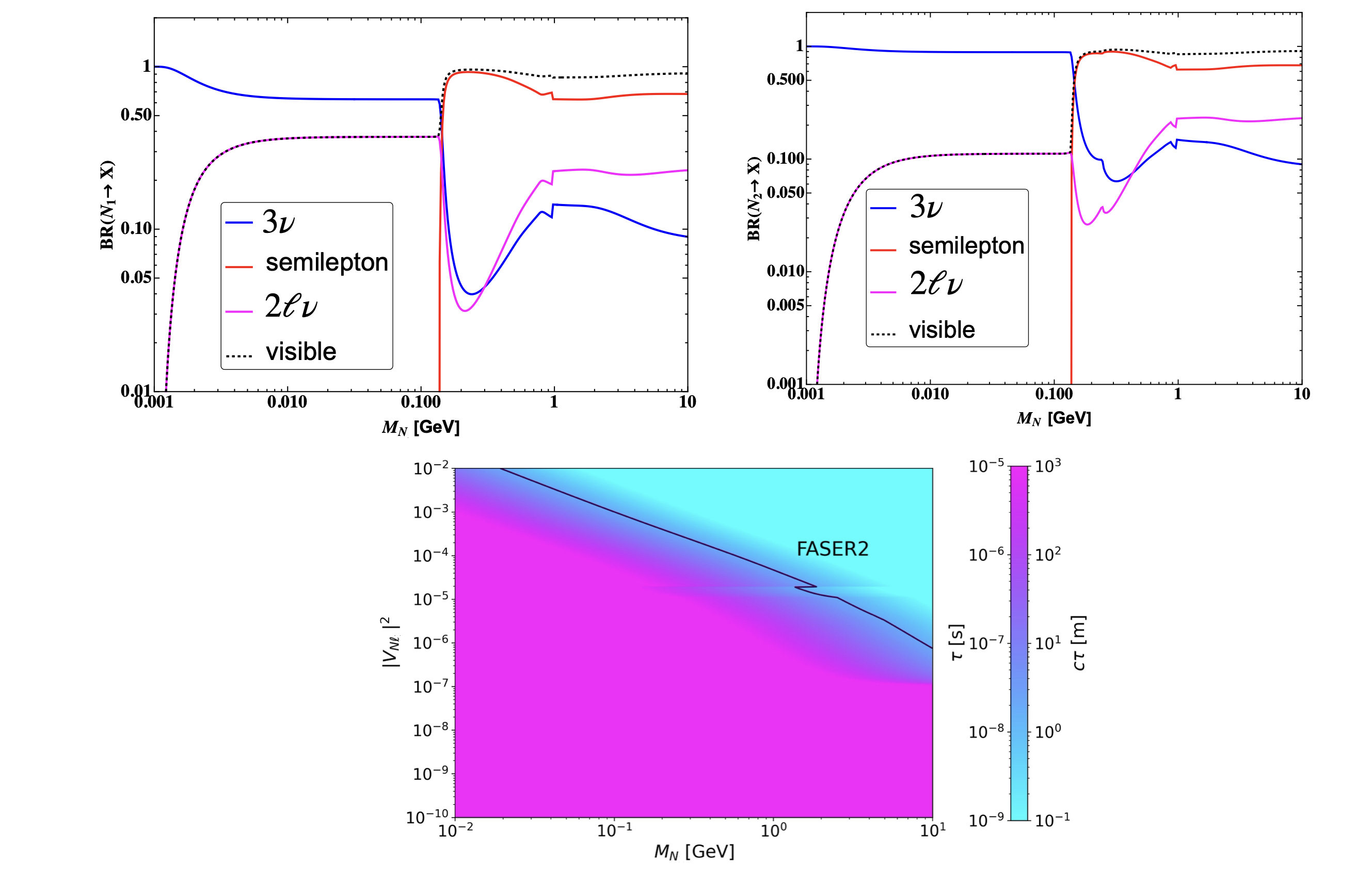}
\caption{Branching ratio of RHNs in different modes (top panel) and its decay length (bottom panel) considering decays of RHN at the rest frame. The dashed black line in each of the upper panel stands for the branching ratio of the RHN to the visible mode. The black contour in the bottom panel stands for the decay length $L_{N}=620$ m relevant for FASER2 detector. }
\label{branching-RHN}
\end{figure}
The partial decay widths of RHN into charged and neutral vector mesons $(V^{+,0})$ are given as
\begin{align}
    \Gamma(N\to \ell^-_1 V^+) &= |V_{\ell_1 N}|^2\frac{G_F^2}{16\pi}M_N^3f_V^2|V_V|^2F_V(y_{\ell},y_V)~, 
    \label{l1v} \\
    \Gamma(N\to \nu_{\ell_1}V^0) &= |V_{\ell_1 N}|^2\frac{G_F^2}{2\pi}M_N^3f_V^2\kappa_V^2 F_V(y_{\nu_\ell},y_V)~,
    \label{nuv}
\end{align}
where $f_V$ is the decay constant of the vector meson $V$, $\kappa_V$ is the corresponding NC coupling of vector meson, defined as,
\begin{align}
\kappa_{\rho^{0}}=\frac{1}{\sqrt{2}}\left(\frac{1}{2}-\text{sin}^{2}\theta_{W}\right),\,\, \kappa_{\omega}=-\frac{1}{3\sqrt{2}}\text{sin}^{2}\theta_{W}\,\, \text{and  } \kappa_{K^{*0}}=\frac{1}{2}\left(\frac{2}{3} \text{sin}^{2}\theta_{W} - \frac{1}{2} \right).
\end{align}
RHNs also decay into opposite sign different flavored quarks through $W$ boson, and the corresponding partial decay widths 
\bea
    \Gamma(N\to \ell_1^- q\bar{q}^\prime)=N_C |V_{\ell_1 N}|^2 |V_{q\bar{q}^\prime}|^2 \frac{G_F^2}{16\pi^3}M_N^5 I_1(y_{\nu_{\ell_1}},y_q, y_{q'})~,
    \label{l1ud}
\eea
where $V_{q\bar{q}^\prime}$ is the corresponding elements of the CKM matrix as listed in Tab.~\ref{tab:mesdec}.  The Similar interaction could be possible through a $Z$ boson exchange, and the corresponding partial decay width is given by 
\bea
    \Gamma(N\to \nu_{\ell_1}q\bar{q}) =
    N_C |V_{\ell_1 N}|^2\frac{G_F^2}{8\pi^3}M_N^5\Big[2(g_L^q g_R^q)I_2(y_{\nu_{\ell}},y_q, y_q)+((g_L^q)^2+(g_R^q)^2)I_1(y_{\nu_{\ell}},y_q, y_q)\Big]~,
    \label{nuqq}
\eea
with $g_L^u = 1/2 - (2/3) \sin^2\theta_{\rm W}$, $g_R^u = - (2/3) \sin^2\theta_{\rm W}$, $g_L^d = -1/2 + (1/3) \sin^2\theta_{\rm W}$, and $g_R^d = (1/3) \sin^2\theta_{\rm W}$.
\begin{table}[]
    \centering
    \begin{tabular}{|c|c|c|c||c|c|c|c|}\hline
       Psudoscalar(P)  & $m_P$ (MeV) & $f_P$(MeV) & $V_P$ & Vector(V)  & $m_V$ (MeV) & $f_V$(MeV) & $V_V$ \\ \hline
        $\pi^\pm$ & 139.6 & 130.7 & $V_{ud}$ & $\rho^\pm$ & 775.8 & 220 & $V_{ud}$ \\ \hline
        $K^\pm$ & 493.7 & 159.8 & $V_{us}$ & $K^{*\pm}$ & 891.66 & 217 & $V_{us}$ \\ \hline
        $\eta$ & 547.8 & 164.7 & -- & $\omega$ & 782.59 & 195 & -- \\ \hline
        $\pi^0$ & 135 & 130 & -- & $\rho^0$ & 776 & 220 & -- \\ \hline
        $K^0$ & 497.6 & 159 & -- & $K^{*0}$ & 896.1 & 217  & -- \\ \hline
    \end{tabular}
    \caption{Masses and decay constants of pseudoscalar and vector mesons following \cite{ParticleDataGroup:2010dbb,CLEO:2005jsh,MILC:2002lnl,Ivanov:2006ni,Feldmann:1999uf,Cvetic:2004qg,Ebert:2006hj}.}
    \label{tab:mesdec}
\end{table}
In the above formulae of the partial decay widths of RHNs, the kinematical functions are give by
\begin{align}
    I_1(x,y,z) &= 
    \int_{(x+y)^2}^{(1-z)^2}\frac{ds}{s}(s-x^2-y^2)(1+z^2-s) \lambda(s,x^2,y^2) \lambda(1,s,z^2)~, \\
    I_2(x,y,z) &=
    yz \int_{(y+z)^2}^{(1-x)^2}\frac{ds}{s}(1+x^2-s) \lambda(s,y^2,z^2) \lambda(1,s,x^2)~, \\
    F_P(x,y) &=
    \lambda(1,x^2,y^2)[(1+x^2)(1+x^2-y^2)-4x^2]~, \\
    F_V(x,y) &=
    \lambda(1,x^2,y^2)[(1-x^2)^2+(1+x^2)y^2-2y^4]~,
\end{align}
with the Callen function
\begin{equation}
    \lambda(x,y,z) = \sqrt{x^2+y^2+z^2-2xy-2yz-2zx}~.
\end{equation}
Since RHNs are Majorana fermions in our scenario, they can decay into the charge conjugates. 
As a result Eqs.~\eqref{nul1l2}, \eqref{l1p}, \eqref{l1v}, and \eqref{l1ud} will contribute to the total decay width in the conjugate form, basically doubling the corresponding contributions, and on the other hand, these contributions will not appear in case of a Dirac fermion. 
Adding all these modes, the total decay width of Majorana RHNs is given as  
\small
\begin{equation}
    \Gamma_N = \sum_{\ell_1, \ell_2 (\ell_1\neq \ell_2)}\left( 2\Gamma(N \to \ell_1^- \ell_2^+\nu_{\ell_2}) + \Gamma(N \to \nu_{\ell_1} \ell_2^- \ell_2^+) \right)  + \sum_{\ell_2} \Gamma(N \to \nu_{\ell_2} \ell_2^- \ell_2^+) + \sum_{\ell_1}\Gamma(N \to \nu_{\ell_1}\nu\bar{\nu})+\Gamma^{\rm semilepton}.
  \label{decn}
\end{equation}
\normalsize
The visible decay of the RHN  will involve all the above modes and conjugates except for the $3\nu$ mode given in Eq.~\eqref{3nu} which is invisible decay mode of the RHNs. Finally, we calculate the semileptonic contribution from the RHNs in inclusive way following 
\begin{equation}
\begin{aligned}
\Gamma^{\rm semilepton} =  
& \theta(\mu_0-M_N)\sum_{\ell_1, P, V}\Big[2\Gamma(N \to \ell_1^-P^+) + 2\Gamma(N \to \ell_1^-V^+) + & \Gamma(N \to \nu_{\ell_1}P^0) + \Gamma(N \to \nu_{\ell_1}V^0)\Big]\\
& + \theta(M_N - \mu_0) \sum_{\ell_1,q,q'}[\Gamma(N \to \nu_{\ell_1} q\bar{q}) + 2\Gamma(N \to \ell_1^- q\bar{q'})]~.  &
\end{aligned}
\end{equation}
Here, $\mu_0$ is the mass threshold for which we consider semileptonic contributions via quark-antiquark production, and we consider $\mu_0=957.8$ MeV below which RHNs decay into light mesons like $\pi^{0, \pm}$, $\rho^{0,\pm}$, $\eta$, $\omega$, and strange mesons like $K^{0, \pm}$ and $K^{*0,\pm}$ being kinematically allowed. 
For mass range $M_N > \mu_0$, we can consider quarks as degrees of freedom, and then the semileptonic decay will consist of quark pairs like $q\bar{q}$ and $q\bar{q}^\prime$. 
\par We show the branching ratios of the RHNs in the upper panel of Fig.~\ref{branching-RHN} where we assume that the first (second) generation RHN dominantly couples to electron (muon). In our scenario the third generation RHN could be potential DM candidate and does not participate in the neutrino mass generation mechanism. With our assumption that the heavy neutrino only mixes with one lepton generation, the branching ratio does not depend on neutrino mixing. The branching ratio of the RHN into $3\nu$ mode dominates over the other modes up to the pion mass, and branching ratio of the RHN into semileptonic decay mode increases above the pion mass threshold. The branching ratio of the $N_{i=1,2}$ decay into visible final states is illustrated by the black dashed line in the left panel of Fig.~\ref{branching-RHN}. This is important for estimating the potential signal yield, as experiments are more likely to detect visible final states rather than the invisible $3\nu$ final states. We find that, once the $N_i$ is heavier than light hadrons, e.g., $\pi$ and $\eta$, $\text{BR}(N_i \to \text{visible}) > 90\%$. Finally, using the total decay width of the RHN from Eq.~(\ref{decn}) in the expression of decay length given by Eq.~(\ref{DL}), we can calculate the corresponding decay length of the RHN assuming RHN decays in rest frame. In the lower panel of  Fig.~\ref{branching-RHN}, we show the decay length of $N_i$ as a function of mass and mixing. This decay length will have same properties for $N_1$ and $N_2$. The branching ratio of the RHNs depend on the mass thresholds of the leptons being manifested in the estimations of branching ratios and decay lengths. For sufficiently small mixing the heavy neutrinos become long-lived particles being capable of traveling macroscopic distances before decaying, resulting in displaced vertex signatures. However for relatively large mass and mixing, heavy neutrino decay-length can be small, so that they could decay promptly. Although here we consider only the parameter space where RHNs are long-lived.
\section{Long-lived particle search experiments}
\label{sec:experiment}
FASER~\cite{Kling:2018wct,FASER:2018ceo,FASER:2018bac,Feng:2017vli,FASER:2018eoc,FASER:2018bac} is a new experiment at the LHC designed to search for light and very weakly-interacting new particles produced in the ATLAS Interaction Point (IP). The detector is located 480 meters from the ATLAS IP, along the beam collision axis line-of-sight (LOS), within the unused service tunnel TI12~\cite{FASER:2022hcn}. However, the overall size of FASER, and therefore its possible decay volume, has been heavily constrained, by the available space underground~\cite{Feng:2022inv}. This provides strong motivation for a larger detector, FASER2, which was already considered in the original FASER proposal~\cite{Feng:2017uoz}. FASER2 involves several important design considerations which were ultimately narrowed down to two preferred choices: (1) a newly constructed, purpose-built facility located approximately 617–682 meters west of the ATLAS IP, and (2) alcoves extending from the existing UJ12 cavern, situated 480–521 meters east of the ATLAS IP \cite{Feng:2022inv,Anchordoqui:2021ghd} for details. For our analysis, we consider the first scenario with the following detector specifications,
\begin{align}
\text{\textbf{FASER2}}:L =620~ \text{m}, \Delta = 10~\text{m},  R = 1~\text{m},  \mathcal{L} = \text{3~ab}^{-1}.
\end{align}
where $L$ is the distance between IP and FASER2 detector, $\Delta$ and $R$ are the depth and radius of the cylindrical shaped detector. The collision energy is assumed to be 14 TeV.  In evaluating the physics reach for the various models discussed below, we will assume that FASER2 can detect all decays of LLPs into visible final states occurring within its decay volume. A schematic depiction of the detector is presented in \cite{FASER:2022hcn}. Finally, we assume that FASER2 will be capable of suppressing potential high-energy backgrounds to a negligible level~\cite{Bohlen:2014buj,Ferrari:2005zk,FASER:2018eoc}. This assumption is justified as this site is shielded from the ATLAS IP by over 200 meters of concrete and rock, making it an ideal location for investigating rare processes and extremely weakly interacting particles or LLPs. It is also worth noting that cosmic ray backgrounds can be effectively distinguished from LLP signals using directionality and timing information. We will also assume a $100\%$ detection efficiency for all visible decay modes.
\begin{figure}[h]
\centering
\includegraphics[scale=0.32]{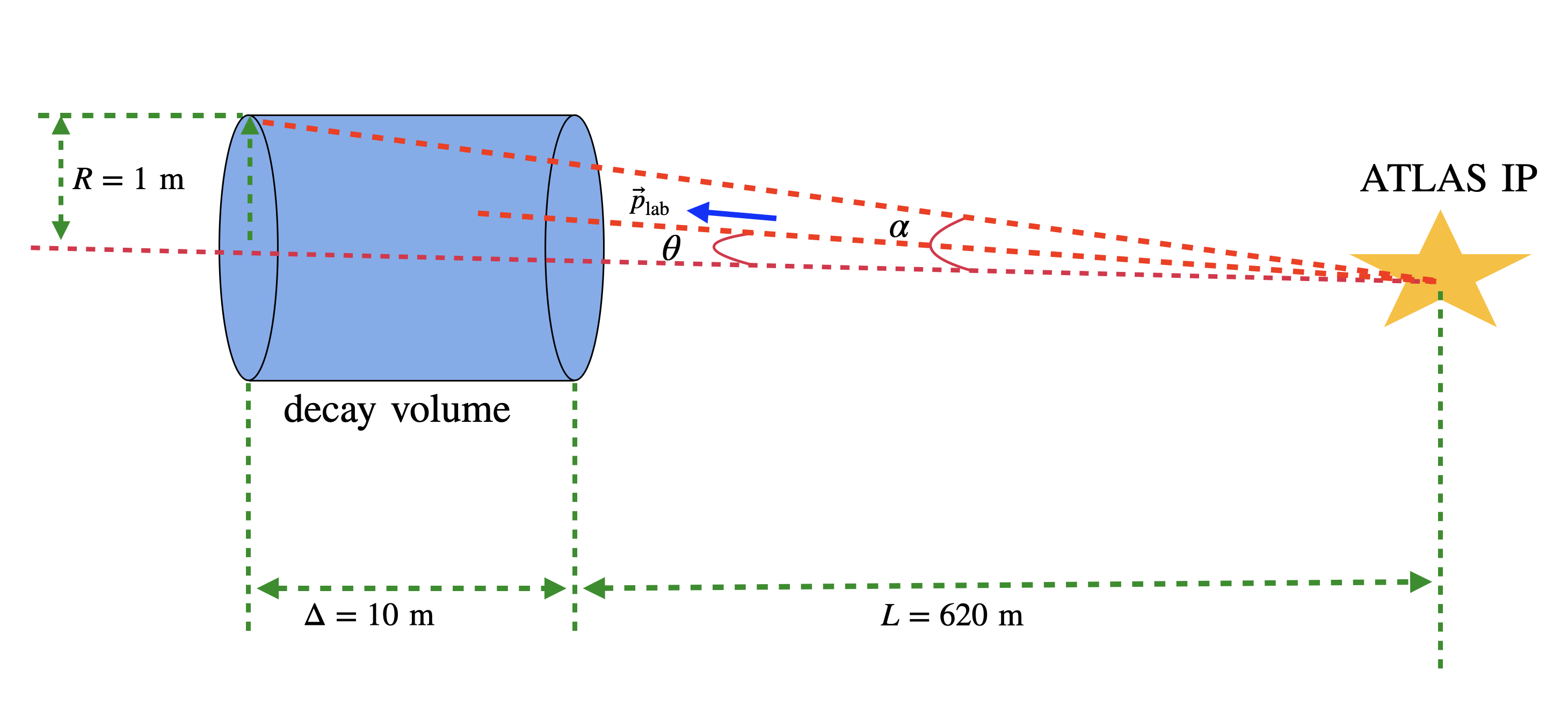}
\caption{Two dimensional schematic representation of FASER2 detector where $\theta$ is the angle of LLPs with the $z$-axis and $\alpha$ is that at the edge of detector where $\tan\theta=\Vec{p}^\perp_\text{lab}/\Vec{p}^{\,z}_\text{lab}$ and $\tan\alpha=R/(\Delta+L)$, respectively. LLPs can reach at the detector if $\theta\leq\alpha$, i.e., $\tan\theta\leq\tan\alpha$.}
\label{fig:llp}
\end{figure}

As shown in Fig.~\ref{fig:llp}, LLPs are typically emitted from the source with a non-zero angle~($\theta$) with respect to the beam axis, only a certain fraction of them reach the detector~($\theta\le \alpha$), and even a smaller fraction decay within the detector. For a LLP of mass $m_{\rm LLP}$ produced at the IP with momentum $p=|\vec{p}_{\rm lab}|$ and angle $\theta$ relative to the beam axis, the probability that it decays inside the FASER2 detector volume is~\footnote{At the LHC, light particles are generally produced with a transverse momentum on the order of their mass, $p_T\sim m_{\rm LLP}$. As a result, LLPs that fall within FASER’s narrow angular acceptance ($\theta\lesssim 1$ mrad) typically carry very high energies, around the TeV scale~\cite{FASER:2018eoc}.}:
\begin{align}
    \mathcal{P}(p, \theta)=(e^{-(L-\Delta)/d_{\rm lab}}-e^{-L/d_{\rm lab}})\Theta(R-\tan\theta L)\approx \frac{\Delta}{d_{\rm lab}}e^{-L/d_{\rm lab}}\,\Theta(R-\theta L).
\end{align}
The first term in the brackets represents the probability that the LLP decays within the interval ($L-\Delta$, $L$), where $d_\text{lab}$ is the decay length of the LLP in the lab frame, and it can be expressed as $d_\text{lab} = c\tau p/m_{\rm LLP}$, with $\tau$ is the LLP’s lifetime. The angular acceptance is determined by the factor if the LLPs emit from the ATLAS point with an angle $\theta$, that is less than the angle made by the edges of the detector with the ATLAS point, $\alpha$ which can be expressed by the above Heaviside function, see Fig.~\ref{fig:llp}. Hence, the fraction of produced LLPs which decays inside the detector volume is evaluated by the acceptance Acc(LLP, $p,\, \theta$) and can be expressed as,
\begin{equation}
    \text{Acc}(\text{LLP}, p, \theta) = \mathcal{P}(p,\theta)~\text{BR}({\,\text{LLP}}\to \text{visible})~,
    \label{fig:angcut}
\end{equation}
Given the Acceptance factor, the number of signal events is then formulated as follows:
\begin{figure}[h]
\centering
\includegraphics[scale=0.45]{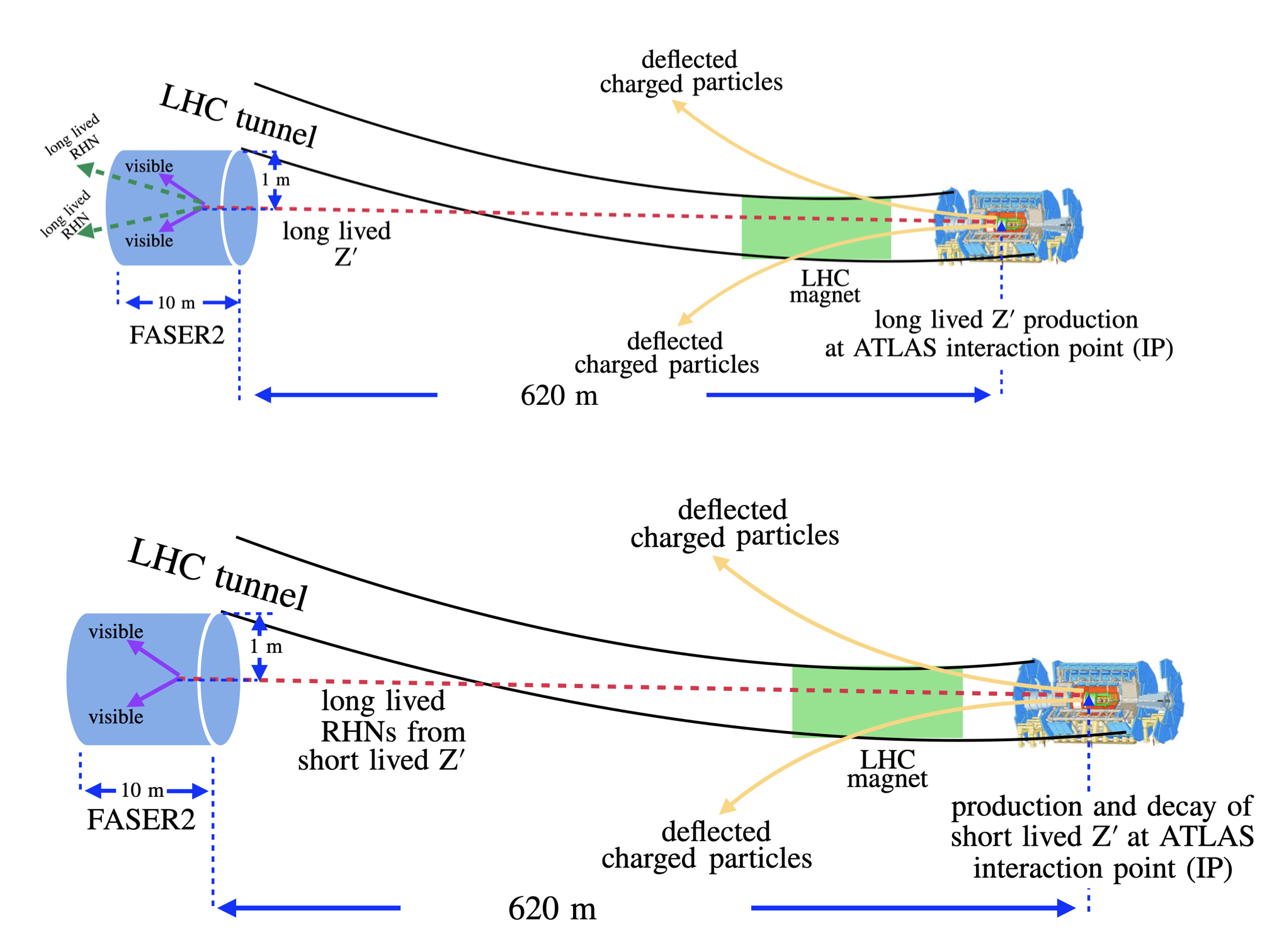}
\caption{Schematic representation of long lived $Z^\prime$ production and decay into visible and long-lived RHN where RHNs are escaping FASER2 detector (upper panel). Production and decay of short-lived $Z^\prime$ at the ATLAS interaction point (IP) and its decay into long-lived RHNs where RHNs decay into visible modes inside the FASER2 detector (lower panel). We used these specifications in this paper.}
\label{fig:llp1}
\end{figure}
\begin{align}
    N_S=\mathcal{L}\int dp d\theta \frac{d\sigma_{pp\to\text{LLP}+X}}{dpd\theta} \times  \text{Acc}(\text{LLP}, p, \theta).
\label{eq:Ns}
\end{align}
There is also a trigger requirement on the LLPs, such that $p>100$~ GeV which is typically fulfilled for LLPs traveling close to the beam collision axis and sufficiently boosted to decay in FASER2 detector.

In this paper, we consider long-lived and short-lived $Z^\prime$ bosons which decay into visible final states~(leptons/hadrons) or pair of RHNs if kinematically accessible. At the ATLAS IP, huge amount of pseudosclar mesons for example,  $\pi^0$ and $\eta$ mesons, are produced by proton-proton collision. The $Z^\prime$ boson can be produced by the decay of these mesons, and the mass of produced $Z^\prime$ is limited by the mass of the meson. In addition, $Z^\prime$ can be produced from proton bremsstrahlung process which is not restricted by kinematic limit of the meson mass. For the bremsstrahlung process, the production cross section is calculated by~\cite{Kim:1973he,Feng:2017uoz,Bauer:2018onh,Asai:2022zxw}
\begin{align}
\label{eq:xsec-pbrems}
    \sigma(p p \to p Z' X) =
    \int \dd p_{Z'}^2 \int \dd \cos\theta_{Z'}^{} \frac{p_{Z'}}{p_{p_i}^{}} w(p_{Z'}^2, \cos\theta_{Z'}^{}) \sigma_{pp}(s')~,
\end{align}
where $\theta_{Z'}$ stands for the angle of $Z'$ respect to the beam axis, $p_{p_i}^{}$ does for the momentum of initial proton in the beam, and $s' = 2 m_p (E_p - E_{Z'}^{})$ with $m_p$ being the proton mass and $E_p~(E_{Z'}^{})$ the energy of initial proton ($Z'$ gauge boson). $\sigma_{pp}(s)$ is the inelastic $pp$ scattering cross-section and the $w(p_{Z'}^2, \cos\theta_{Z'}^{})$ is the splitting function and can be found in \cite{Kim:1973he,Tsai:1973py,Blumlein:2013cua,Asai:2022zxw}. The differential production rate of the $Z'$ gauge boson is then calculated as
\begin{align}
\label{eq:prod-rate_p-brem}
    \frac{\dd N^{\rm brem}}{\dd p_{Z'}^2 \dd \cos\theta_{Z'}^{}} =
    \frac{p_{Z'}^{}}{p_{p_i}} w(p_{Z'}^2, \cos\theta_{Z'}^{}) \frac{\sigma_{pp}(s')}{\sigma_{pp}(s)}~,
\end{align}
with $s = 2 m_p E_{p_i}$. On the other hand, for the  production of $Z'$ from rare decays of mesons, the differential production rate is calculated as
\begin{align}
\label{eq:prod-rate_p-meson}  
    \frac{\dd N^M}{\dd p_M^2 \dd \cos\theta_M^{}} =
    \frac{d\sigma(p p \to M X)}{\dd p_M^2 \dd \cos\theta_M^{}} \cdot {\rm BR}(M \to Z' \gamma)~,
\end{align}
where $p_M^{}$ and $\theta_M^{}$ are the momentum and angle of meson respect to the beam axis, respectively.
Once produced, these $Z^\prime$ gauge bosons could pass through the natural and man-made material and reaches at the FASER2 detector. In our analysis, we consider the visible decay modes of the $Z^\prime$.  Produced $Z^\prime$ can also decay into a pair of RHNs which can further decay into visible modes involving hadrons in the semileptonic and SM charged leptons in the fully leptonic modes. 
As shown in Fig.~\ref{fig:llp1}, for our study, we consider two different scenarios: 
\begin{itemize}
\item[(i)] \textbf{Long-lived $Z'$ + Long-lived $N_i$:} We first consider the scenario where both the $Z^\prime$ and $N_i$ are long-lived. As a result, produced $Z^\prime$ might reach the detector and can decays to visible modes or pair of RHNs, see the upper panel of Fig.~\ref{fig:llp1}. However, as the RHNs are also long lived, they will decay outside the detector. Therefore, in this case, visible signal consists of the visible decay modes of $Z^\prime$ only. The acceptance is then evaluated as
\begin{equation}
\text{Acc}(\text{LLP}, p_{Z'}, \theta_{Z'}) = \mathcal{P}(p_{Z'},\theta_{Z'})~\text{BR}(Z'\to \text{visible})~.
\label{eq:accep_1}
\end{equation}
From Eqs.~(\ref{eq:prod-rate_p-brem}), \eqref{eq:prod-rate_p-meson} and (\ref{eq:accep_1}), the total expected numbers of events in FASER2 is given by
\begin{align}
\label{eq:num_p-brem}
N_{\rm event}^{\rm p\mathchar`-brem}& =
    N_p\, |F_1(m_{Z'}^2)|^2 \int \dd\, p_{Z'}^2 \int \dd \cos\theta_{Z'}^{}\, \frac{\dd N^{\rm brem}}{\dd\, p_{Z'}^2 \dd \cos\theta_{Z'}^{}}\, \Theta(\Lambda_{\rm QCD}^{} - q^2) \cdot {\rm Acc}(\text{LLP},p_{Z'},\theta_{Z'})~,\\ 
\label{eq:num_p-meson}
N_{\rm event}^{\rm p\mathchar`-meson}& =
    N_p\, \sum_{M = \pi^0, \eta} \int \dd p_M^2 \int \dd \cos\theta_M^{} \int \dd p_{Z'}^2 \int \dd \cos\theta_{Z'}^{}\, \frac{\dd N^M}{\dd p_M^2 \dd \cos\theta_M^{}} \cdot {\rm Acc}(\text{LLP},p_{Z'},\theta_{Z'})~,
\end{align}
where $N_p$ stands for the number of protons on target. To estimate the number of signal events using the methods described above, we utilize the FORESEE Python package~\cite{Kling:2021fwx}. After supplying the formulas for $Z'$ production from bremsstrahlung, the package can generate the differential spectra of the $Z'$. In the upper left and upper right  panels of Fig.~\ref{fig:spec1}, we show the distribution of $\pi$ and $\eta$ mesons in the ($\theta$, $p$) plane, where $\theta$ and $p$ are the meson’s angle with respect to the beam axis and their momentum, respectively. The concentration of events in Fig.~\ref{fig:spec1} along the line $p_T=\Lambda_{\rm QCD}\sim$ 0.25 GeV suggests a characteristic momentum transfer scale. Notice, the high multiplicity of mesons with momentum  $p>100$ GeV produced at small angles $\theta< 10^{-3}$.  They serve as efficient sources of forward, high-momentum $Z'$ as shown in the lower panel of Fig.~\ref{fig:spec1}. If these $Z'$ are long-lived, they can reach the detector and decay to visible modes. The bottom left and right panel stands for $x_H=-2$ and $x_H=2$, respectively where we fix the coupling and mass as $g_X=10^{-6}$, $M_{Z'}=0.1$ GeV.
This signature will be effective to rule out the small mass and/or small coupling regions where $Z'$ decay length $L_{Z'}$ is close to the distance between IP and FASER2 detector. 
\begin{figure}[!htbp]
\centering
\includegraphics[scale=0.53]{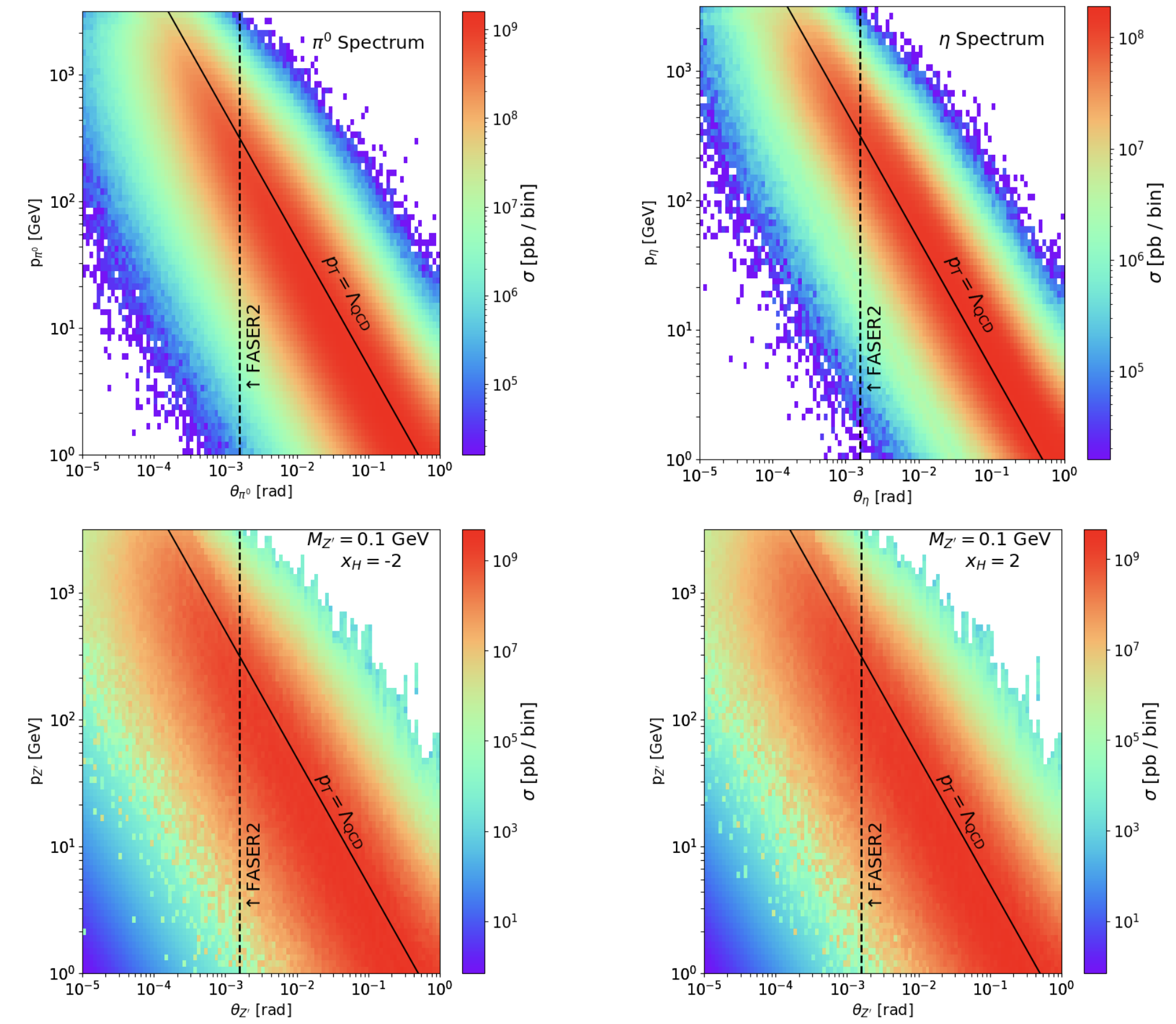}
\caption{Momentum ($p$) versus angle ($\theta$) relative to the beam axis for various particle spectra. The upper left (right) panel shows the $\pi^0(\eta)$ meson spectrum produced from the $pp$ collision at LHC. Distributions of $Z'$ bosons dominantly produced from meson decays, with parameters $M_{Z'} = 0.1$ GeV (normalized by $g_X^2$) in the long-lived $Z^\prime$ scenario with $x_H = -2~(2)$ are shown in the lower left (right) panel. The color bar indicates the differential cross section per bin (in pb). These distributions illustrate the kinematic properties of the particles as they are produced in the forward region at the LHC.}
\label{fig:spec1}
\end{figure}
\item[(ii)] \textbf{Short-lived $Z'$ + Long-lived $N_i$:} Now we consider the scenario of short-lived $Z'$ and long-lived RHNs. As a result, $Z^\prime$ promptly decays into either visible modes or a pair of RHNs right after the production, as shown in the bottom panel of Fig.~\ref{fig:llp1}. If these RHNs are long-lived, they can travel to the detector to decay into visible modes contributing to the visible signals. Then, the acceptance is evaluated as
\begin{align}
\label{eq:accep_2}
\text{Acc}({\rm LLP}, p_i^1, \theta_i^1, p_i^2, \theta_i^2) = \text{BR}(Z' \to 2N_i) \Big[ &\text{Acc}(N^1_i, p_i^1, \theta_i^1)   + \text{Acc}(N_i^2, p_i^2, \theta_i^2)  \Big]~  \text{  with   } \vec{p}_{Z'} = \vec{p}_i{}^1+\vec{p}_i{}^2,
\end{align}
where $N_i^1$ and $N_i^2$ denote first and second RHNs from RHN pair of same generation $i$, with momentum $p_i^{1}$ and $p_i^2$, respectively. Further the acceptance for RHNs $\text{Acc}(N_i^\alpha, p_i^\alpha, \theta_i^\alpha)$ can be expressed as,
\begin{align}
\text{Acc}(N_i^\alpha, p_i^\alpha, \theta_i^\alpha) = \mathcal{P}(p_i^\alpha,\theta_i^\alpha)~\text{BR}(N_i\to \text{visible})~,
\end{align}
The kinematic distribution of the $Z'$ will exhibit a similar pattern to the previous case, with a few differences. This signal is only relevant if $Z'$ is short-lived which is true if the coupling and/or mass is relatively large. If the mass of $Z'$ is large, the productions are dominated by bremsstrahlung process. As $Z'$ decays two identical RHNs, $p_i^1,p_i^2\equiv p_{N_i} \approx p_{Z'}/2$ and $\theta_i^1,\theta_i^2\equiv \theta_{N_i}\approx \theta_{Z'}$ since $p_{N_i}\gg M_{N_i}$ in the forward direction. Therefore, the distributions of $p_{N_i}$ and $\theta_{N_i}$ for the $N_i$ should closely resemble those for the $Z'$. This information allows us to rewrite the acceptance factor as follows
\begin{align}
\label{eq:accep_22}
\text{Acc}({\rm LLP}, p_i^1, \theta_i^1, p_i^2, \theta_i^2)\equiv \text{Acc}({\rm LLP},p_{N_i},\theta_{N_i}) \approx 2\,\text{BR}(Z' \to 2N_i) \text{Acc}(N_i, p_{N_i}, \theta_{N_i})~,
\end{align}
with $p_{N_i}\approx p_{Z'}/2$ and $\theta_{N_i}\approx \theta_{Z'}$. Then the expected numbers of events from bremsstrahlung process and mesons decay in FASER2 are given by,
\begin{align}
\label{eq:num_p-brem-1}
N_{\rm event}^{\rm p\mathchar`-brem}& =
    N_p\, |F_1(m_{Z'}^2)|^2 \int \dd\, p_{Z'}^2 \int \dd \cos\theta_{Z'}^{}\, \frac{\dd N^{\rm brem}}{\dd\, p_{Z'}^2 \dd \cos\theta_{Z'}^{}}\, \Theta(\Lambda_{\rm QCD}^{} - q^2) \cdot \sum_{i}{\rm Acc}(\text{LLP},p_{N_i},\theta_{N_i})~,\\ 
\label{eq:num_p-meson-1}
    N_{\rm event}^{\rm p\mathchar`-meson} &=
    N_p\, \sum_{M = \pi^0, \eta} \int \dd p_M^2 \int \dd \cos\theta_M^{} \int \dd p_{Z'}^2 \int \dd \cos\theta_{Z'}^{}\, \frac{\dd N^M}{\dd p_M^2 \dd \cos\theta_M^{}} \cdot \sum_{i}{\rm Acc}(\text{LLP},p_{N_i},\theta_{N_i})~,
\end{align}
where $N_p$ stands for the number of protons on target. This signatures will be effective to rule out large coupling~($g_X$) and/or large $Z'$ mass, as well as small mixing $(V_{\ell N})$ and/or small RHN mass regions.
\end{itemize}
\section{Results and discussion}
\label{sec:RD}
In this section, we show the expected sensitivity of FASER2 for our two considered scenarios: long-lived and short-lived $Z'$. In both the cases we fix $M_N=M_{Z'}/3$ so that $Z'\to NN$ is kinematically allowed. As the $Z'$ and $N$ can decay into pairs of visible particles, their decay vertex can be reconstructed in the FASER2 detector. Also as the FASER2 site is shielded from the ATLAS IP by over 200 meters of concrete and rock, we safely ignore the background. Hence we only require $N_S=N_{\rm event}^{\rm p\mathchar`-brem}+N_{\rm event}^{\rm p\mathchar`-meson}>3$ to define the sensitivity at $95\%$ confidence level. In the following we discuss the sensitivities in the $g_X-M_{Z'}$ plane and $|V_{\ell N}|^2-M_N$ for different values of $x_H$. 
\begin{figure}[h]
\centering
\includegraphics[scale=0.465]{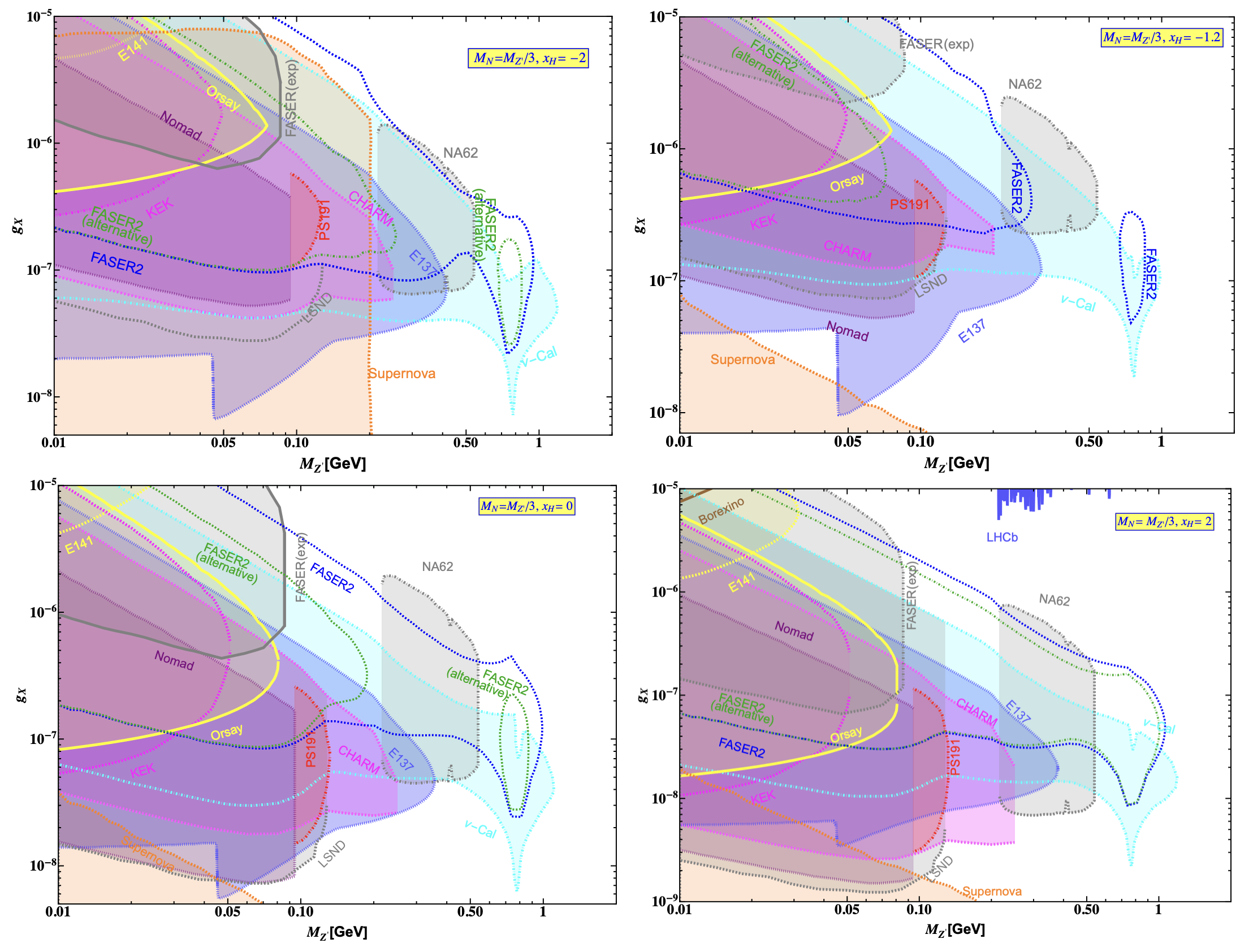}
\caption{Limits on general $U(1)$ couplings as a function of $M_{Z^\prime}$ shown for long-lived $Z^\prime$ scenarios following long-lived RHNs for different $x_H$ considering $|V_{\ell N}|^2= 10^{-6}$ and $M_{N_{1,2}}= M_{Z^\prime}/3$. In each panel, blue dashed and darker green dashed lines stand for the $U(1)_X$ (FASER2) and `alternative' (FASER2 (alternative)) scenarios. Shaded regions are already ruled out by various existing experimental results.}
\label{fig:llp-Zp}
\end{figure}
\subsection{Long-lived $Z'$+long-lived RHNs}
In this case long-lived $Z^\prime$ decays into visible final states at the detector and visible decays of long-lived RHNs are undetected. Note that we are considering long-lived RHNs which are controlled basically by two parameters, namely $M_N$ and $|V_{\ell N}|^2$. The heavy neutrino decay length in the rest frame as a function of its mass and the light-heavy mixing parameter is shown in Fig.~\ref{branching-RHN}. Following this, we fix the mixing $|V_{\ell N}|^2=10^{-6}$, so that RHNs are long-lived and hence undetected. 
\par We show the sensitivities on $g_X-M_{Z^\prime}$ plane for long-lived $Z^\prime$ boson taking $x_H = -2$, $-1.2$, $0$ and $2$ in Fig.~\ref{fig:llp-Zp}. In each panel, blue dashed and darker green dashed lines stand for the general U$(1)_X$ case and `alternative' scenarios. We see from Figs.~\ref{branching-Zp} and \ref{branching-Zp-1} that in the `alternative' case, BR$(Z^\prime \to \text{visible})$ is smaller than that in the general $U(1)_X$ case due to the difference in the $U(1)_X$ charges of the RHNs. Therefore the predicted exclusion region on $g_X-M_{Z^\prime}$ plane is wider for the general $U(1)_X$ case compared to the `alternative' case. Also we find that the depending on the values of $x_H$, the exclusion region changes. This is due to the following two reasons: depending on the values of $x_H$, the quark coupling with $Z'$ varies and hence the $Z'$ production cross-section; $Z'$ decay length/branching ratio to visible mode also varies with $x_H$ due to different strength of $Z'$ coupling with quarks and leptons. For example, bounds on the coupling corresponding to the charges $x_H=-1.2$ and $x_H=2$, could be considered as weakest and strongest, respectively. For $x_H=-1.2$, the branching ratio $\text{BR}(Z'\to NN)$ maximizes~(see Fig.~\ref{branching-Zp-1}), hence the weakest bound. We also see that irrespective of $x_H$ charges, the best sensitivity on the coupling lies within $10^{-7} \leq g_X \leq 10^{-5}$ when $M_{Z'}$ lies between $0.1$ GeV $\leq M_{Z^\prime} \leq 1$ GeV. For small masses and couplings, the sensitivity is the same for $U(1)_X$ and the `alternative' case, despite $\mathrm{BR}(Z'\!\to\!\mathrm{visible})$ being larger in $U(1)_X$. In the `alternative' case, different RHN charge assignments enhance the probability of $Z'$ decaying inside the detector at small $g_X$.
\par Finally, we compare our parameter regions with those obtained from different electron beam-dump experiments involving Orsay \cite{Davier:1989wz}, KEK \cite{Beer:1986qr}, E141 \cite{Riordan:1987aw}, E137 \cite{Bjorken:1988as}, E774 \cite{Bross:1989mp} and GEMMA \cite{Beda:2009kx,Lindner:2018kjo}; proton beam-dump experiments involving $\nu$-Cal \cite{Blumlein:2011mv,Blumlein:2013cua}, LSND \cite{LSND:1997vqj}, PS191 \cite{Bernardi:1985ny}, NOMAD \cite{NOMAD:2001eyx}, and CHARM \cite{CHARM:1985anb}. We find that our study can exclude parts of the parameter space which is not ruled out by existing bounds obtained from different beam-dump experiments. This exclusion regions crucially depends on the information about the decay of long-lived $Z^\prime$ into a pair of long-lived RHNs. The exclusion region in $g_X-M_{Z^\prime}$ plane will also depend on whether $Z^\prime \to N N$ mode is kinematically allowed~\cite{Asai:2022zxw}.  
\subsection{Short-lived $Z'$+long-lived RHNs}

\begin{figure}[h]
\centering
\includegraphics[scale=0.435]{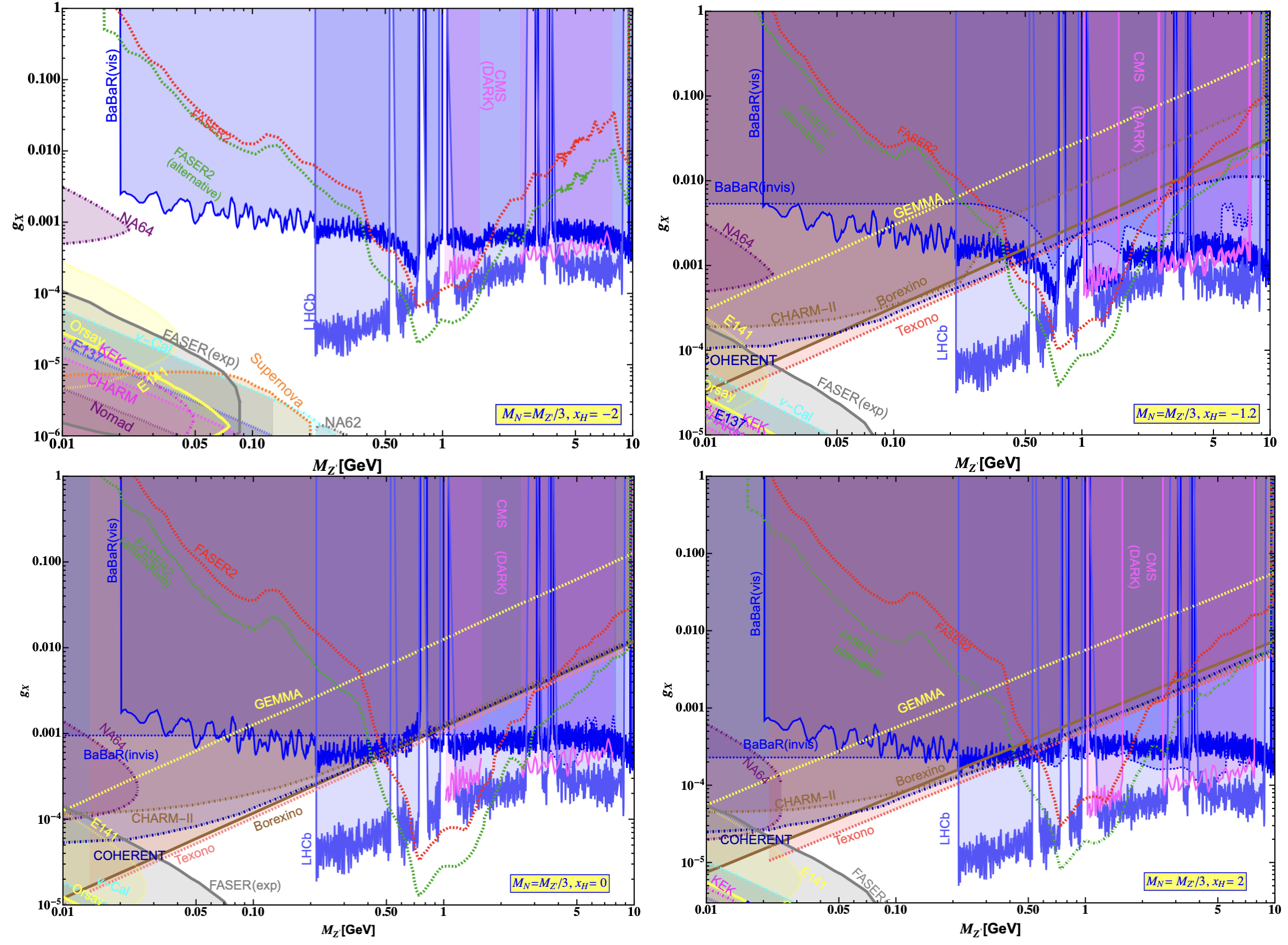}
\caption{Limits on general $U(1)$ couplings as a function of $M_{Z^\prime}$ for short-lived $Z^\prime$ scenarios following the decay of long-lived RHNs for the general $U(1)_X$ (red dashed) and `alternative' (darker green dashed) cases considering $|V_{\ell N}|^2= 10^{-6}$ and $M_{N_{1,2}}= M_{Z^\prime}/3$. Shaded regions are already ruled out by existing limits from different experiments.}
\label{fig:slp-Zp}
\end{figure}

Apart from long-lived $Z^\prime$ scenario, we consider that $Z^\prime$ could be short-lived and RHNs produced from $Z^\prime$ will manifest long-lived nature and decay inside the detector. Taking visible final states from the RHNs into account we estimate bounds on $g_X-M_{Z^\prime}$ plane for $x_H=-2$, $-1.2$, $0$ and $2$ for the general $U(1)_X$ and `alternative' cases in Fig.~\ref{fig:slp-Zp}. Again in these analyses we considered light-heavy neutrino mixing squared $|V_{\ell N}|^2 \simeq \mathcal{O}(10^{-6})$ to have long-lived RHNs  where $\ell=e,~\mu$ following $M_{N_{1,2}}= M_{Z^\prime}/3$.  Considering $N \to e/\mu+\rm{(associated~particles)}$ and $N \to \nu_{e/\mu}+ \rm{(associated~particles)}$ modes we obtain bounds in general $U(1)_X$ (red dashed) and `alternative' (darker green dashed) cases where associated particles produced from different decay modes of the RHNs are mentioned previously following the corresponding branching rations shown  in Fig.~\ref{branching-RHN}. We performed the same analysis considering muon instead of electrons and found that the limits are extremely close with those with the electrons supporting the fact that in our model $Z^\prime$ interacts equally with different generations of the fermions, in other words flavor universal.
\begin{figure}[h]
\centering
\includegraphics[scale=0.44]{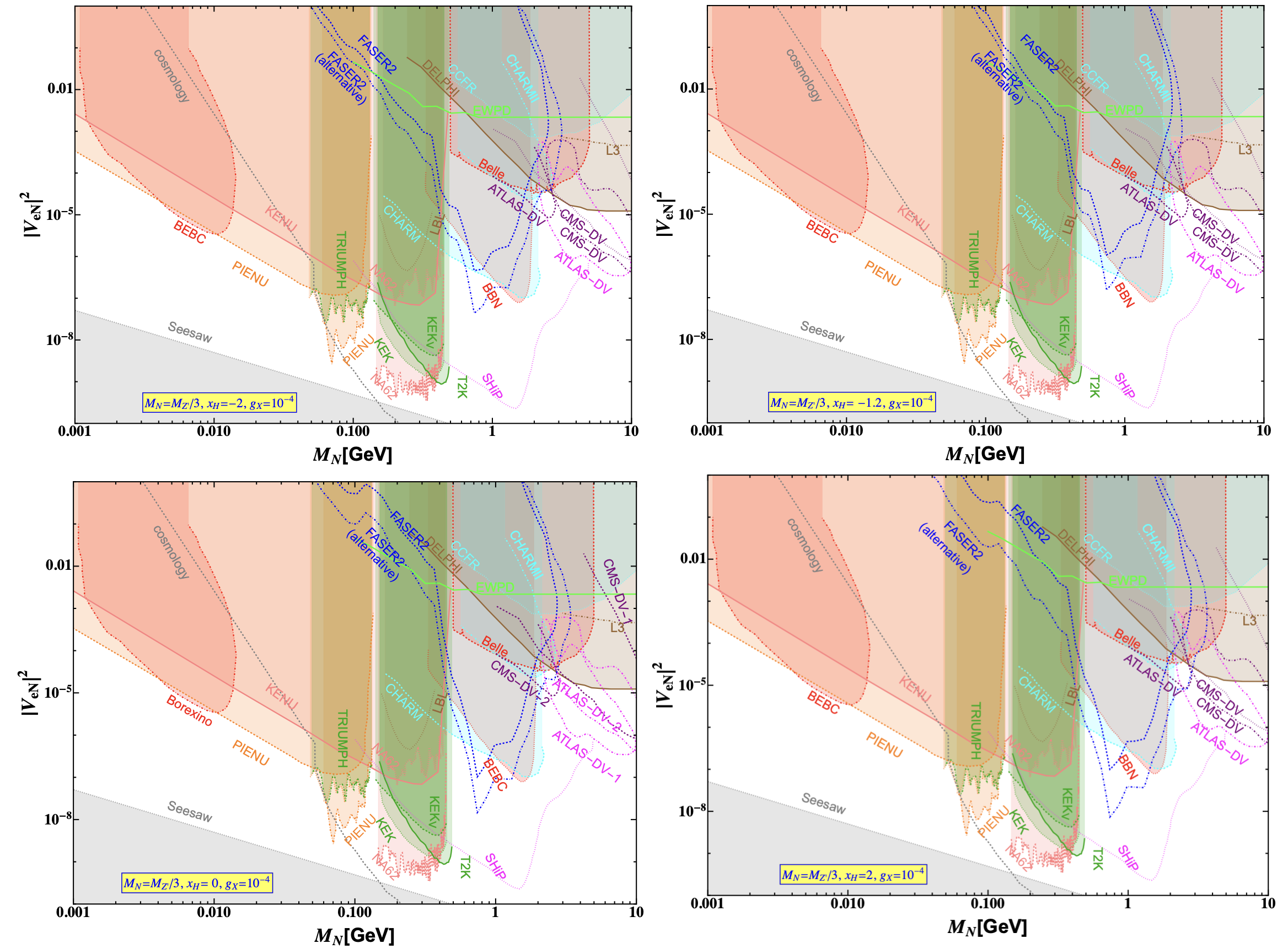}
\caption{Limits on light-heavy neutrino mixing as a function of $M_N$ considering $g_X^{} =10^{-4}$ and $M_{N}= M_{Z^\prime}/3$ as benchmarks such that short-lived $Z^\prime$ produced at ATLAS IP can decay into long-lived RHNs into electron and associated states. Blue dotted and blue dot-dashed lines stand for $U(1)_X$ and `alternative' scenarios, respectively. Shaded region is already ruled out from a variety of existing experiments.}
\label{fig:slp-Zp-e}
\end{figure}
\begin{figure}[h]
\centering
\includegraphics[scale=0.44]{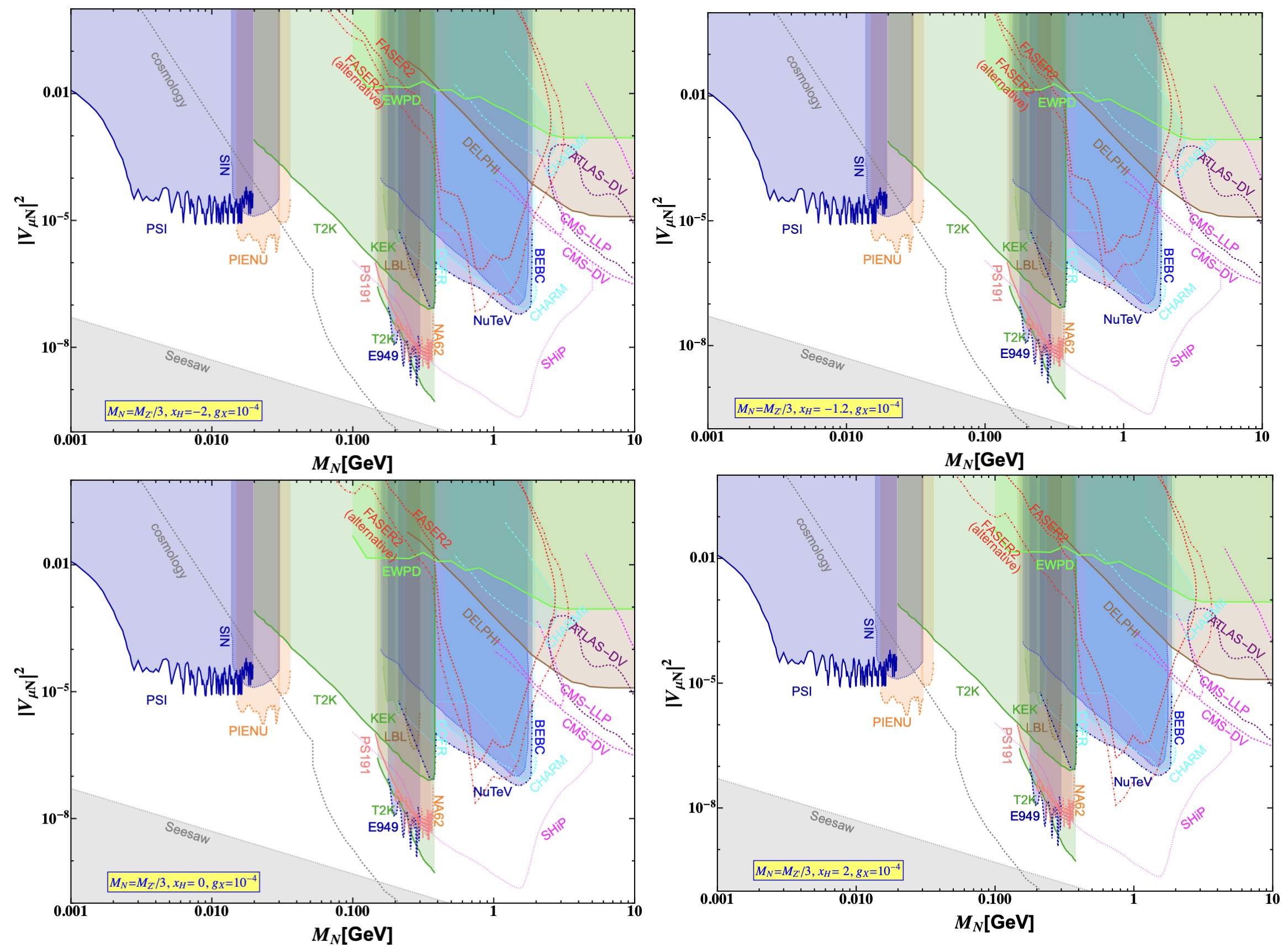}
\caption{Limits on light-heavy neutrino mixing as a function of $M_N$ considering $g_X^{} =10^{-4}$ and $M_{N}= M_{Z^\prime}/3$ as benchmarks such that short-lived $Z^\prime$ produced at ATLAS IP can decay into long-lived RHNs into muon and associated states. Red dotted and red dot-dashed lines stand for $U(1)_X$ and `alternative' scenarios, respectively. Shaded region is already ruled out from a variety of existing experiments.}
\label{fig:slp-Zp-m}
\end{figure}
We obtain stronger bounds from the `alternative' scenario due to the larger $Z'\to NN$ branching ratio compare to the $U(1)_X$ case. We find that strongest bound in short-lived $Z^\prime$ scenario $g_X\sim 10^{-4 (5)}$ could be observed when 0.6 GeV $\leq M_{Z^\prime} \leq 2.1$ GeV in the general $U(1)_X$ (`alternative') case. We compare our results with the bounds obtained on $g_X-M_{Z^\prime}$ plane from different scattering experiments involving neutrino-electron scattering measurement from TEXONO \cite{TEXONO:2009knm}, BOREXINO \cite{Borexino:2008gab,Bellini:2011rx}, GEMMA \cite{Beda:2010hk}, and CHARM-II \cite{CHARM-II:1994dzw}, Neutrino-nucleon scattering measurement from COHERENT~\cite{COHERENT:2020ybo} and dark photon search experiments like LHCb \cite{LHCb:2019vmc}, BaBaR \cite{BaBar:2014zli,BaBar:2017tiz}, and CMS \cite{CMS:2023slr}. The shaded parameter regions are ruled out by these experiments are shown in Fig.~\ref{fig:slp-Zp} for different $x_H$ taken from \cite{Asai:2023mzl,KA:2023dyz}.

In general, for given $x_H$, the number of signal events is determined by four free parameters: $g_X$, $M_{Z'}$, $M_N$ and $|V_{\ell N}|^2$. Hence, to explore the sensitivity in the parameter space $M_N-|V_{\ell N}|^2$, it is necessary to fix the values of $g_X$ and $M_{Z'}$. To study this sensitivity in the case of short-lived $Z^\prime$, we consider $g_X=10^{-4}$ as a benchmark taking $M_N=M_{Z^\prime}/3$. In Figs.~\ref{fig:slp-Zp-e} and \ref{fig:slp-Zp-m}, we show the limits on the light-heavy neutrino mixing angles for different values of $x_H$. We considered  $N \to e (\mu)+\rm{(associated~particles)}$ mode to obtain limits on $|V_{e(\mu) N}|^2$. The blue (red) dashed and blue (red) dotdashed lines in each panel stands for $U(1)_X$ and `alternative' scenario, respectively for the electron (muon) case. We find that the bound on the mixing angle is always tighter for the 'alternative' scenario compared to the general $U(1)_X$ case. 

Projected limits on $|V_{e N}|^2-M_N$ plane, shown in Fig.~\ref{fig:slp-Zp-e}, from FASER2 are strong in $U(1)_X$ case within $0.7$ GeV $\leq M_N \leq 0.9$ GeV for $x_H=0$ and $2$. We find that projected bounds on $|V_{e N}|^2-M_N$ plane obtained from FASER2 stay mostly within the shaded regions except for $0.5$ GeV $\leq M_{N} \leq 1.8$ GeV depending on $x_H$ for the `alternative' scenario.  Strongest limits on light-heavy neutrino mixing could reach around $\mathcal{O}(10^{-7})$ for the $U(1)_X$ case and $\mathcal{O}(10^{-8})$ in `alternative' case, respectively when $M_{Z^\prime} \simeq 0.6$ GeV. We find that strong limit on the light-heavy mixing for $x_H=-1.2$ could be obtained around $10^{-7}$ from the `alternative' case whereas bounds obtained from the $U(1)_X$ case are weak staying within the shaded region. Shaded regions are already ruled out by different existing experimental bounds.

Projected limits on $|V_{\mu N}|^2-M_N$ plane, shown in Fig.~\ref{fig:slp-Zp-m}, from FASER2 are strong around $M_{Z^\prime}=0.75$ GeV for $U(1)_X$ case for $x_H=0$ and $2$. Prospective bounds are stronger in the `alternative' case compared to the $U(1)_X$ scenario for all $x_H$. Strongest bounds on light-heavy neutrino mixing could reach around $10^{-7}$ for the $U(1)_X$ case and $10^{-8}$ for the `alternative' case. For $x_H=-1.2$, prospective bounds on the light-heavy mixing from the `alternative' case could be as strong as $5\times 10^{-7}$ for a very narrow region around $M_{Z^\prime}=0.75$ GeV and those obtained from the $U(1)_X$ case stay within the shaded region. Looking at the `alternative' scenario we find that $0.5$ GeV $\leq M_{N} \leq 1.5$ GeV could be an interesting range of RHN mass to obtain a strong bound on the light-heavy neutrino mixing and that shrinks to $0.7$ GeV $\leq M_N \leq 0.9$ GeV for the $U(1)_X$ case for $x_H=0$ and $2$. These results will depend on the choice and relation between $M_{Z'}$ and $M_N$. 

Existing bounds on $|V_{e(\mu) N}|^2-M_N$ plane, shown in Figs. \ref{fig:slp-Zp-e} and \ref{fig:slp-Zp-m}, are obtained from different experiments and have been discussed in  \cite{Chun:2019nwi,Bolton:2019pcu,Fernandez-Martinez:2023phj} previously. Strongest bound on $|V_{eN}|^2-M_N$ plane are obtained peak search experiments at TRIUMPH \cite{Britton:1992xv} and PIENU \cite{PIENU:2017wbj,Bryman:2019bjg,PIENU:2019usb} from $\pi^+ \to e^+ N$ decay within $10^{-3}$ GeV $\leq M_N \leq 0.15$ GeV where limits from Borexino \cite{Borexino:2013bot} were strongest within $0.003$ GeV $\leq M_N \leq 0.016$ GeV following $N\to \nu e^+ e^-$. Dominant contribution on the limits on the light-heavy mixing comes from $K^+ \to e^+ N$ decay studied by NA62 \cite{NA62:2020mcv,NA62:2025csa} within 0.15 GeV $\leq M_N \leq 0.45$ GeV. In neutrino experiments like T2K \cite{T2K:2019jwa} searched for RHN manifest in-flight decay through CC interaction and detected in near detector providing a competitively strong limit on the mixing within $0.4$ GeV $\leq M_N \leq 0.5$ GeV. RHN heavier than kaon could be studied at  BEBC \cite{WA66:1985mfx,Barouki:2022bkt} and CHARM \cite{CHARM:1985nku} experiments within $0.5$ GeV $\leq M_N \leq 2$ GeV and $2$ GeV $\leq M_N \leq 2.25$ GeV, respectively. In addition to that studies from Belle \cite{Zhou:2021ylt}, DELPHI \cite{DELPHI:1996qcc}, ATLAS \cite{Tastet:2021vwp,ATLAS:2022atq} and CMS \cite{CMS:2022fut} experiments provide strong bounds on the light-heavy mixing within $2.25$ GeV $\leq M_N \leq 2.5$ GeV, $2.5$ GeV $\leq M_N \leq 2.75$ GeV, $2.75$ GeV $\leq M_N \leq 15$ GeV and $15$ GeV $\leq M_N \leq 16$ GeV, respectively. In addition to that we show recent bounds NA62 \cite{NA62:2020mcv} by green dot-dashed line, displaced vertex searches from ATLAS \cite{ATLAS:2019kpx,ATLAS:2022atq,ATLAS:2024fdw} by dashed cyan line, CMS \cite{CMS:2024xdq,CMS:2024bni} by dotted and dot-dashed cyan lines and prospective limits from SHiP \cite{Alekhin:2015byh,SHiP:2018xqw} by green dashed line, respectively. In the context of $|V_{\mu N}|^2-M_N$ bounds we find that PSI \cite{Daum:1987bg} provides strong limit $1.2\times 10^{-3}$ GeV $\leq M_N \leq 0.0175$ GeV where RHN is produced from pion decay. PIENU \cite{PIENU:2019usb} provide strong limit within $0.0175$ GeV $\leq M_N \leq 0.025$ GeV. We obtain bounds on the light-heavy mixing from T2K \cite{Arguelles:2021dqn} when RHN is above pion mass, MicroBooNe \cite{Kelly:2021xbv},  and KEK \cite{Hayano:1982wu} experiments within the mass range $0.0325$ GeV $\leq M_N \leq 0.375$ GeV providing a strong bound on the mixing. We also find limits on the mixing between 0.2 GeV $\leq M_N \leq 0.3$ GeV from the E949 experiment at BNL \cite{BNL-E949:2009dza}, $0.375$ GeV $\leq M_N \leq 2.0$ GeV following BEBC \cite{WA66:1985mfx}, NuTeV \cite{NuTeV:1999kej}, CHARM \cite{CHARM:1985nku} and 2.0 GeV $\leq M_N \leq 11$ GeV from ATLAS \cite{ATLAS:2019kpx,ATLAS:2022atq} and CMS \cite{CMS:2022fut} experiments. We show the NA62 bounds on the light-heavy neutrino mixing angle by light-red dot and dot-dashed lines from \cite{NA62:2025csa}. Prospective bounds on the light-heavy mixing from the SHiP experiment are shown by the dotted magenta line \cite{SHiP:2018xqw}. Combined limits from cosmology from CMB and BBN are taken from \cite{Vincent:2014rja,Dolgov:2000pj,Ruchayskiy:2012si,Gelmini:2020ekg,Langhoff:2022bij,Sabti:2020yrt,Boyarsky:2020dzc} providing bounds up to $M_N \leq 0.2$ GeV on $|V_{e(\mu) N}|^2-M_N$ plane in Figs.~\ref{fig:slp-Zp-e} and \ref{fig:slp-Zp-m}  by gray dot-dashed line. Using the seesaw relation using neutrino mass, for example, $m_\nu \leq 0.1$ eV we estimate the theoretical lower bound on the mixing square $|V_{e(\mu) N}|^2 \leq 10^{-12} (100~{\rm GeV)} / M_N$ shown by `Seesaw' in gray dashed line in Figs.~\ref{fig:slp-Zp-e} and \ref{fig:slp-Zp-m}, respectively. 
\section{Conclusions}
\label{sec:conc}
We studied two types of $U(1)_X$ extensions which we dubbed as general $U(1)_X$ and `alternative' scenarios where we introduced three generations of RHNs which could generate light neutrino mass through seesaw mechanism after U(1)$_X$ symmetry breaking. In these scenarios a neutral BSM gauge boson $Z^\prime$ is evolved and its mass is also governed by the VEV of the $U(1)_X$. In this work, we explored the projected sensitivities of FASER2 detector at the LHC to the $Z'$ and RHNs. In both the models we considered $Z'$ couples with SM fermions with different strength depending on the choice of $x_H$. Hence depending on the mass, $Z^\prime$ can be produced from proton-bremsstrahlung and rare meson decays at the ATLAS IP. After the production $Z'$ can decay into visible final states (leptons/hadrons) or pair of RHNs if kinematically accessible. We considered two type of scenarios: (i) long-lived $Z'$ + long-lived RHNs and (ii) short-lived $Z'$ + long-lived RHNs to estimate bounds on the model parameters like $U(1)_X$ gauge coupling,  $Z^\prime$ mass, light-heavy neutrino mixing and RHN mass. In the case of long-lived $Z'$, they can travel to the FASER2 detector and considering the visible decay modes of the $Z'$, we estimate the bounds on $g_X-M_{Z'}$ plane. We find that the exclusion region depends on the values of the $x_H$ and the best sensitivity on the coupling could be obtained around $10^{-5} \leq g_X \leq 10^{-6}$ for $M_{Z'}\simeq \mathcal{O}(100)$ MeV and $10^{-6} \leq g_X \leq 10^{-7}$ for $0.2$ GeV $\leq M_{Z^\prime} \leq 0.8$ GeV. For long-lived $Z^\prime$ $U(1)_X$ could provide stronger bounds compared to the existing experimental results for $0.08$ GeV $\leq M_{Z^{\prime}}\leq 0.8$ GeV irrespective of charges, however, for the `alternative' case such strong bounds compared to the existing experimental results could be obtained for $x_H=0$ and $2$. On the other hand, in case of short-lived $Z'$ scenario, we consider that it can decay to RHNs at the ATLAS IP and long-lived RHNs can travel a distance up to the FASER2 detector where they could decay into visible final states. 
We first estimated bounds on the $g_X-M_{Z'}$ plane considering the mixing $|V_{\ell N}|^2\simeq 10^{-6}$ to make sure that RHNs are long-lived. We found that the FASER2 experiment could probe $M_{Z^\prime} \simeq \mathcal{O}(1)$ GeV within $10^{-5} \leq g_X \leq 10^{-4}$ depending on $U(1)$ scenarios and their charges. Finally, in this context we estimated prospective bounds on the light-heavy neutrino mixing taking $g_X=10^{-4}$ in our account. We found that depending on general $U(1)$ scenario and their charges FASER2 could provide strong prospective bound on the light-heavy neutrino mixing within $10^{-7} \leq |V_{\ell N}|^2 \leq 10^{-8}$ for 0.5 GeV $ \leq M_N \leq 1.8$ GeV depending on the general $U(1)$ scenarios and corresponding $U(1)$ charges which could be probed in future.
\black
\begin{acknowledgements}
S.K.A is supported by JST SPRING, Grant Number JPMJSP2119. The work of S.M. is supported by KIAS Individual Grants (PG086002) at Korea Institute for Advanced Study. 
\end{acknowledgements}
\bibliographystyle{utphys}
\bibliography{bibliography}
\end{document}